\documentclass{aa}  

\usepackage{graphicx}
\usepackage{txfonts}
\usepackage{lipsum}
\usepackage{subcaption}         
\usepackage{lscape}             
\usepackage{placeins}           
                                
\usepackage[dvipsnames]{xcolor}
\usepackage[colorlinks=true, linkcolor=blue, citecolor=blue, urlcolor=blue]{hyperref}
\usepackage{float}
\usepackage{amsmath}
\usepackage{needspace}
\usepackage[font=small,labelfont=bf]{caption}
\usepackage{dblfloatfix}
\begin{document}

\titlerunning{CAPOS abundances in NGC 6304}
\authorrunning{C. Montecinos et al.}

\title{CAPOS: The bulge Cluster APOgee Survey}
\subtitle{XI. Unraveling the chemical composition of the bulge globular cluster NGC 6304}

\author{Carolina Montecinos\inst{1}
        \and Doug Geisler\inst{1,2}
        \and Cesar Mu\~{n}oz\inst{1}
        \and Sandro Villanova\inst{3}
       }

\institute{Departamento de Astronom\'ia, Facultad de Ciencias, Universidad de La Serena,
              Raúl Bitrán 1305, La Serena, Chile\\
              \email{carolina.montecinos@userena.cl}
         \and
             Departamento de Astronom\'ia, Casilla 160-C, Universidad de Concepci\'on,
             Concepci\'on, Chile
         \and
             Universidad Andres Bello, Facultad de Ciencias Exactas, Departamento de F{\'i}sica y Astronom{\'i}a - Instituto de Astrof{\'i}sica, Autopista Concepci\'on-Talcahuano 7100, Talcahuano, Chile}      

   \date{Received October 3, 2025; accepted November 17, 2025}

 
  \abstract
  {Bulge globular clusters are key to understanding the formation and chemical evolution of the ancient central component of our Galaxy. Thanks to CAPOS, the bulge Cluster APOgee Survey, we can mitigate the observational difficulties limiting access to these objects in the optical and investigate them in more detail in the near-IR.}
   {Our goal is to perform a rigorous abundance analysis on a large number of member stars in the metal-rich bulge globular cluster NGC 6304, using high-resolution spectra, in order to determine its detailed chemical composition and study the relationship of this globular cluster with its halo counterparts. In addition, we investigate chemical patterns that allow us to identify multiple populations.}
   {This analysis is based on spectroscopic data obtained by CAPOS, which uses the APOGEE-2S spectrograph to observe in the H band. The abundances of 17 elements (C, N, O, Na, Mg, Al, Si, S, K, Ca, Ti, V, Cr, Mn, Fe, Ni, and Ce) were derived using the BACCHUS code, using atmospheric parameters from both APOGEE's ASPCAP pipeline and those derived independently from photometry (\textit{Gaia} and 2MASS) to cross validate our results with the ASPCAP values and provide a consistent chemical analysis.}
   {We derived a mean iron abundance of $[{\rm Fe/H}] = -0.45 \pm 0.05$ using the ASPCAP stellar parameters, and $[{\rm Fe/H}] = -0.45 \pm 0.08$ when using photometric stellar parameters, with no evidence of an intrinsic metallicity spread. NGC 6304 shows a typical enhancement in $\alpha$ elements, with $[\alpha/{\rm Fe}]{\rm spec} = +0.24 \pm 0.07$ and $[\alpha/{\rm Fe}]{\rm phot} = +0.23 \pm 0.08$, similar to what is observed in other globular clusters. We find a significant spread in [N/Fe], with $\sigma_{\rm spec} = 0.54$ and $\sigma_{\rm phot}  = 0.46$, along with a clear C–N anticorrelation. Furthermore, we detect a correlation of Ce with both N and Al, consistent with patterns observed in some metal-rich bulge globular clusters but not all.}
   {Our study provides the first comprehensive spectroscopic evidence for multiple populations in NGC 6304. We also find a significant star-to-star variation in Na, but a minimal variation in O, in concordance with trends found in other metal-rich bulge clusters. The absence of the Mg-Al anticorrelation supports the evidence that the MgAl cycle is not active in globular clusters at high metallicity. The observed correlation between Ce and both N and Al suggests that the enrichment of these elements may be driven by asymptotic giant branch stars, positioning Ce as an element involved in the multiple population phenomenon in metal-rich globular clusters. We find generally that abundances are consistent with those of bulge field stars of similar metallicity, suggesting a similar origin and chemical evolution.}

   \maketitle
%

\section{Introduction}

Our understanding of globular clusters (GCs) experienced a paradigm shift after observations revealed that stars within a GC exhibit star-to-star variations in certain chemical abundances, known today as the multiple population (MP) phenomenon (\citealp{2004ARA&A..42..385G}; \citealp{2009A&A...505..117C};  \citealp{2017A&A...601A.112P}; \citealp{2018ARA&A..56...83B}; \citealp{2019A&A...622A.191M}; \citealp{2020MNRAS.492.1641M}). All observed Galactic GCs above a certain mass (M $\sim$ 5$\times10^{4}$M$_{\odot}$) and age limit ($\sim$2 Gyr) present the MP phenomenon (\citealp{2018ARA&A..56...83B}), with Ruprecht 106 being the only confirmed exception found until now among classical, old, massive GCs (\citealp{2013ApJ...778..186V}; \citealp{2021MNRAS.503..867F}). 
These abundance variations are observed in elements involved in proton-capture reactions during hydrogen burning at high temperatures, caused by the CNO, NeNa, and MgAl cycles (\citealp{1993PASP..105..301L}). The affected elements are light, alpha, and odd-Z elements (C, N, O, Mg, Na, and Al), producing the Carbon-Nitrogen (C-N), Sodium-Oxygen (Na-O), and Magnesium-Aluminum (Mg-Al) anticorrelations, respectively. These chemical patterns are the famous signatures of the presence of MP and are observed in almost all GCs, especially the Na-O anticorrelation (\citealp{2009A&A...505..117C}; \citealp{2011A&A...535A..31V};  \citealp{2015AJ....149..153M}; \citealp{2018A&A...620A..96M}, \citealp{2017A&A...601A.112P};  \citealp{2024MNRAS.528.1393S}). More recently, variations have also been detected even in the neutron-capture element Cerium (\citealp{2021ApJ...918L...9F}), at least in some GCs, but not all (e.g., \citealp{Uribe2025}).

The most widely accepted scenario for the origin of MP is the self-enrichment hypothesis, where a second population (2P) forms from material polluted by the ejecta of evolved first population (1P) stars (\citealp{2004ARA&A..42..385G}; \citealp{2007A&A...470..179P}; \citealp{2010A&A...516A..55C}; \citealp{2011A&A...533A.120V}). The nature of the polluting stars remains uncertain, but several candidates have been proposed, including massive asymptotic giant branch (AGB) stars (\citealp{2016ApJ...831L..17V}; \citealp{2016MNRAS.458.2122D}; \citealp{2018MNRAS.475.3098D}), fast-rotating massive main-sequence (MS) stars (\citealp{2007A&A...470..179P}; \citealp{2007A&A...464.1029D}; \citealp{2013A&A...552A.121K}), massive MS binary stars (\citealp{2009A&A...507L...1D}; \citealp{2013MmSAI..84..171I}), and supermassive stars (\citealp{2014MNRAS.437L..21D}).

Bulge GCs (BGCs) are considered an independent system (\citealp{1995AJ....109.1663M}) that formed in situ, in contrast to halo GCs, which were mostly accreted (\citealp{2019A&A...630L...4M}). However, their study has been significantly limited because of the high extinction and severe crowding toward the central regions of the Milky Way (MW; \citealp{2012A&A...543A..13G}; \citealp{2020A&A...644A.140S}). Nevertheless, despite these observational difficulties, several studies have been carried out to catalog the BGC population. For instance, \citet{2016PASA...33...28B} identified 42 GCs in the Galactic bulge (GB); in addition, \citet{2024A&A...687A.201B} expanded their catalog by adding new GC candidates, increasing the BGC population to as much as 61, with a metallicity distribution characterized by two peaks at [Fe/H] $\sim$ -1.1 and -0.5 dex. More recently, \citet{G2025} performed a quantitative assessment aimed at ensuring a genuine BGC sample by cross-referencing multiple independent studies including \citet{2019A&A...630L...4M}, \citet{2020MNRAS.491.3251P}, \citet{2022MNRAS.513.4107C}, \citet{2024MNRAS.528.3198B}, and \citet{2024OJAp....7E..23C} and identified 40 confirmed BGCs, with a very similar metallicity distribution to that found by \citet{2024A&A...687A.201B}. Therefore, the BGC population is extensive and particularly important, as they represent the ancient in situ bulge population and are thus key tracers of the early stages of formation and chemical evolution of the oldest stellar populations that formed in situ in the MW's main progenitor.

In particular, some studies of MP in metal-rich BGCs have found that they present a significant star-to-star spread in Na, but little or no variation in O (\citealp{2017A&A...605A..12M, 2018A&A...620A..96M, 2020MNRAS.492.3742M}; \citealp{2021MNRAS.503.4336M}), contrary to the typical Na-O anticorrelation observed in halo GCs (\citealp{2010A&A...516A..55C}). Given that BGCs include the highest metallicity of all MW GCs, detailed chemical analyses of these objects are essential to extending our understanding of the MP phenomenon into the high-metallicity regime. However, this requires high-resolution, high signal-to-noise ratio (S/N) spectroscopy of large samples, which is limited using optical spectroscopy. In this context, high spatial resolution near-IR observations are critical.

The Apache Point Observatory Galactic Evolution Experiment II (APOGEE-2; \citealp{2017AJ....154...94M}), part of the Sloan Digital Sky Survey (SDSS-IV; \citealp{2017AJ....154...28B}), was designed to meet these requirements and has been highly successful. However, the SDSS-IV survey did not prioritize BGCs (\citealp{2020MNRAS.492.1641M}). To address this limitation and complement the relatively small number of BGCs observed by the main APOGEE-2 survey, CAPOS the (bulge Cluster APOgee Survey - \citealp{2021A&A...652A.157G}) was created. CAPOS is part of the SDSS-IV APOGEE-2 survey, using the high-resolution (R $\sim$ 22,500) near-IR spectrograph APOGEE-2S, and observed a total of 18 GCs located within a region of $\pm10^{\circ}$ × $\pm10^{\circ}$ around the Galactic center, most of which have only been poorly studied previously. The main goal of CAPOS is to obtain detailed abundances and kinematics for a significant number of member stars, assembling the most complete sample possible of BGCs. Additionally, CAPOS aims to investigate the presence of MP in BGCs, which has been limited almost exclusively to the halo and disk GCs.

In this context, a number of articles from the CAPOS series have been published (\citealp{2021A&A...652A.157G}; \citealp{2021A&A...652A.158R}; \citealp{2022A&A...658A.116F}; \citealp{2023MNRAS.526.6274G}; \citealp{2025A&A...696A.154H}; \citealp{2025A&A...699A.128B}; \citealp{2025A&A...701A.159F}; \citealp{G2025}). All of these studies provide valuable information on the chemical evolution of BGCs, as well as disk GCs. Thanks to CAPOS, it is now possible to investigate key aspects of MP in BGCs through a homogeneous and systematic analysis, something that was previously very difficult to achieve due to observational limitations. 
\begin{figure}
\centering
\includegraphics[width=0.40\textwidth, height=0.27\textheight]{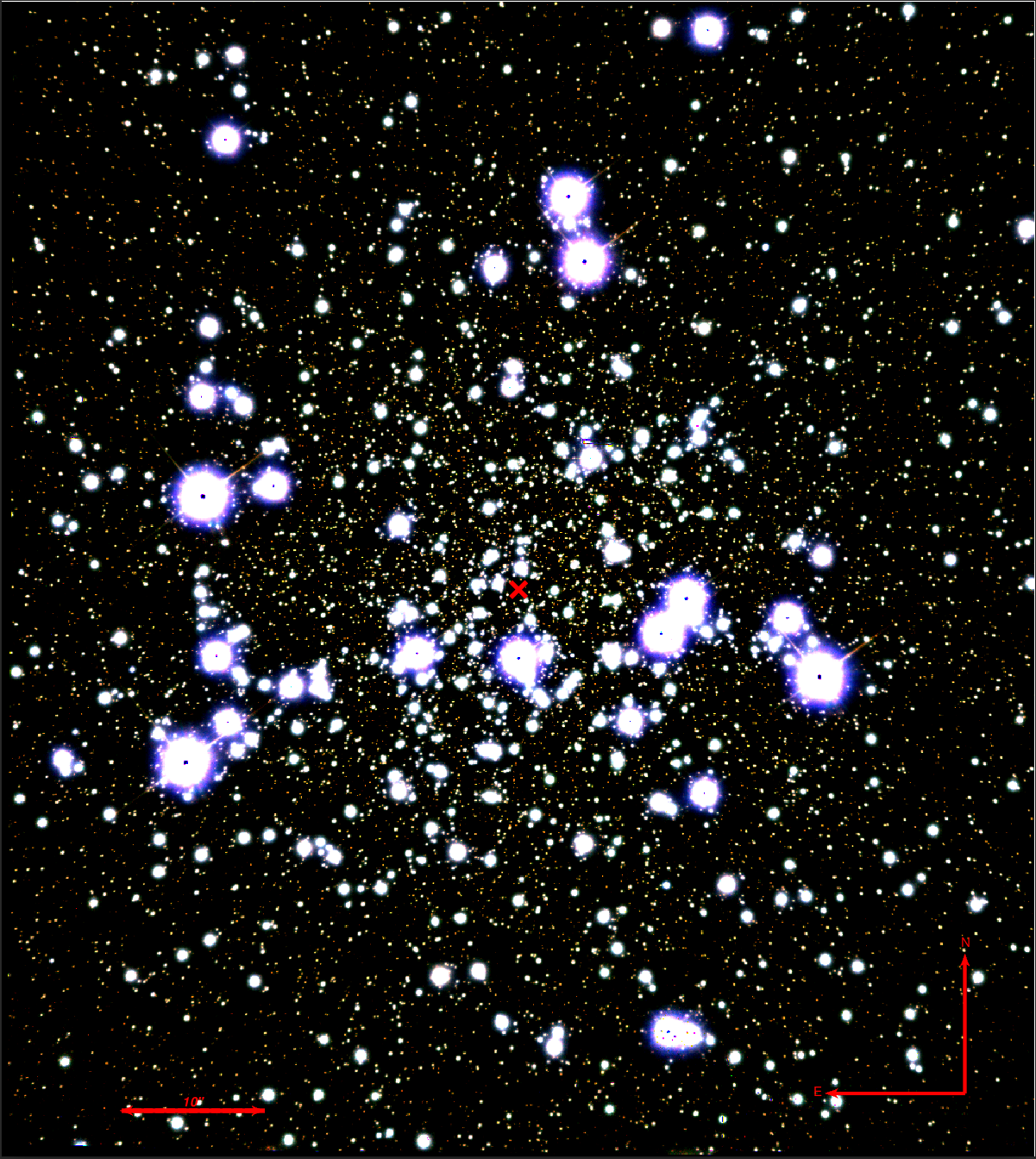}
\caption{Multiband (JHKs-combined color) infrared view of NGC 6304 obtained with the Gemini South Adaptive Optics Imager+Gemini Multi-Conjugate Adaptive Optics System (SAOI+GeMS; Haro et al. in preparation)} 
\label{fig:1}
\end{figure}
In this article, we present for the first time a detailed analysis of chemical abundances and MP of NGC 6304 (Figure \ref{fig:1}), observed as part of CAPOS. NGC 6304 is the most metal-rich GC of the sample, with a metallicity of [Fe/H] = -0.45 (\citealp{1996AJ....112.1487H}). Previous studies of this cluster have been predominantly photometric, yielding a range of metallicity estimates. For example, \citet{2005MNRAS.361..272V} derived [Fe/H] = –0.70 from near-IR photometry, and \citet{2020ApJ...891...37O} obtained [Fe/H] = –0.48 based on HST photometry. This GC has been classified as a bulge globular cluster by several studies (\citealp{2016PASA...33...28B}; \citealp{2019A&A...630L...4M}; \citealp{2020MNRAS.491.3251P}; \citealp{2022MNRAS.513.4107C}; \citealp{G2025}) and as an in situ GC by \citet{2024MNRAS.528.3198B} and \citet{2024OJAp....7E..23C}. According to the \href{https://people.smp.uq.edu.au/HolgerBaumgardt/globular/}{Baumgardt database}, NGC 6304 is located at 2.19$\pm$0.13 kpc from the Galactic center (l = 355.826$^{\circ}$, b = 5.376$^{\circ}$), at 6.15$\pm$0.15 kpc from the Sun and has a mass of 1.03$\pm$0.04 $\times10^{5}$ $M_{\odot}$. This cluster has an age of 12.3 Gyr (\citealp{2020ApJ...891...37O}), an absolute visual magnitude of M$_{V}$ = -7.30 and a color excess of E(B - V) = 0.54 (\citealp{1996AJ....112.1487H}).

This article is organized as follows: In Section 2 we describe the observations and data. In Section 3 we explain the methodology (atmospheric parameters, chemical abundances, errors, and the quality of the lines). In Section 4 we present our results for iron, alpha, and iron-peak elements. In Section 5 we present the analysis of MP and finally in Section 6 we present our concluding remarks.
\section{Observations and data}

\subsection{APOGEE-2 survey}

APOGEE-2 (\citealp{2017AJ....154..198Z};  \citealp{2021AJ....162..302B}; \citealp{2021AJ....162..303S}), part of the SDSS-IV (\citealp{2017AJ....154...28B}), is a high-resolution ($R \sim 22,500$) near-infrared spectroscopic survey that systematically sampled $\sim$ 657,000 giant stars observing the three components of the MW (bulge, disk and halo), providing radial velocities (RVs), stellar parameters, and chemical abundances to study kinematics, morphology, and chemical evolution of our Galaxy. As an extension of the original APOGEE survey (\citealp{2017AJ....154...94M}), APOGEE-2 combined the observations of two 300-fiber APOGEE spectrographs (\citealp{2019PASP..131e5001W}): APOGEE-2N mounted on the 2.5 m Sloan Foundation Telescope (\citealp{2006AJ....131.2332G}) at Apache Point Observatory (New Mexico) and APOGEE-2S mounted on the 2.5 m Ir\'en\'ee du Pont telescope (\citealp{1973ApOpt..12.1430B}) at Las Campanas Observatory (Chile), expanding the scope of the study to the Southern Hemisphere and enabling comprehensive coverage of the Galaxy. Both instruments capture most of the H band between 15,100 and 17,000 {\AA} across three detectors. Small spectral coverage gaps exist between $\sim$ 15,800 - 15,900 {\AA} and $\sim$16,400 - 16,500 {\AA}.

The last APOGEE-2 data catalog was published in SDSS-IV Data Release 17 (DR17) in December 2021 (\citealp{2022ApJS..259...35A}), which is accessible on the SDSS-IV Science Archive Server\footnote{\url{https://www.sdss4.org/dr17/irspec/spectro_data/}}. The data reduction process for DR17 is detailed in \citet{2015AJ....150..173N}. Atmospheric stellar parameters, RVs, and chemical abundances were derived using the APOGEE Stellar Parameters and Chemical Abundance Pipeline ({\ttfamily ASPCAP}; \citealp{2016AJ....151..144G}). {\ttfamily ASPCAP} employs a grid of synthetic spectra generated from {\ttfamily MARCS} model atmospheres (\citealp{2008A&A...486..951G}) to match the observed spectra. The validation of atmospheric stellar parameters and chemical abundances is extensively analyzed in \citet{2018AJ....156..125H} and \citet{2018AJ....156..126J, 2020AJ....160..120J}. The atomic and molecular line list used for synthesizing spectra are described in \citet{2015ApJS..221...24S}, \citet{2016ApJ...833...81H}, \citet{2017ApJ...844..145C}, and \citet{2021AJ....161..254S}.
\begin{figure}
\sidecaption
\includegraphics[width=9cm]{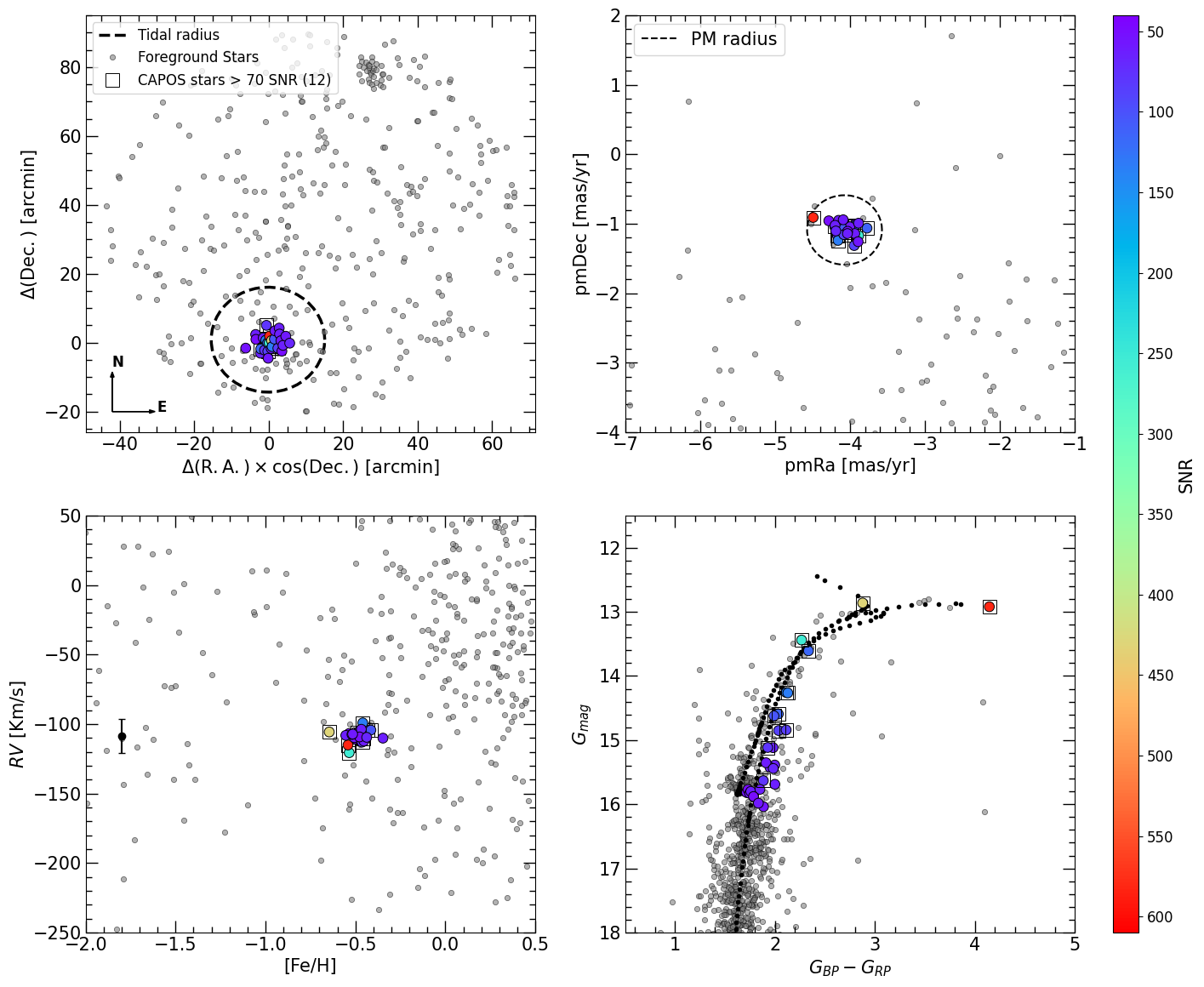}
\caption{Selection of NGC 6304 membership. \textit {Top left}: Spatial distribution of observed stars within the APOGEE-2 survey area. The dashed black circumference indicates the tidal radius of the cluster ($t_{\mathrm{r}}$=15.25\arcmin). NGC 6316 can be seen to the NE. \textit {Top Right}: PM distribution of the stars within the tidal radius obtained from \textit{Gaia} DR3. The dashed circumference has a radius of 0.5 [mas/yr]. \textit{Bottom left}: Selection in the RV vs. [Fe/H] plane from {\ttfamily ASPCAP} for the candidate members. The point with error bars shows the mean RV and its standard deviation from the \citet{2023MNRAS.521.3991B} database. \textit{Bottom right}: G vs. ($G_{\mathrm{bp}}$ - $G_{\mathrm{rp}}$) CMD. Final selected members are shown in all plots as points color-coded according to their S/N, while the field stars are represented as gray dots. The plotted isochrone of 12 Gyr represents the best PARSEC fit. Black squares in the plots are those stars with S/N > 70.}
\label{fig:2}
\end{figure} 

\subsection{NGC 6304}  

NGC 6304 was observed as part of the CAPOS survey (\citealp{2021A&A...652A.157G}) through various CNTAC\footnote{CNTAC:
 Chilean Telescope Allocation Committee} Contributed program allocations (PI D. Geisler). The APOGEE-2S plug-plate containing NGC 6304 was centered at (l,b) $\sim$ (355.83°, 5.38°) and corresponds to the APOGEE field name 356+06-C, shared with NGC 6316 (\citealp{2025A&A...701A.159F}).
 
The membership selection process is illustrated in Figure \ref{fig:2} and is described as follows: First, we selected all stars located within the cluster tidal radius taken from \href{https://people.smp.uq.edu.au/HolgerBaumgardt/globular/}{Baumgardt database}\footnote{\url{https://people.smp.uq.edu.au/HolgerBaumgardt/globular/}} ($t_{\mathrm{r}} = 15'.2$; represented by the dashed black circumference in the top left panel of Figure \ref{fig:2}).  Then, to refine the membership selection, we used the proper motions (PM) from \textit{Gaia} DR3 (\citealp{2023A&A...674A..37G}) and selected stars within a radius of 0.5 mas $yr^{-1}$ from the mean PM value (dashed black circumference of the top right panel of Figure \ref{fig:2}). This radius is optimal for maximizing members and minimizing contamination from field stars (also used in \citealp{G2025}). As additional criteria, the mean RV and metallicity values ([Fe/H]) for the cluster were applied. We have used the values of both RVs and [Fe/H] derived from {\ttfamily ASPCAP}, considering [Fe/H] only for membership selection. Stars with RVs within 2$\sigma$ (-120 to -96 km$s^{-1}$) of the mean value of -108 km$s^{-1}$ were included, while stars with [Fe/H] within 3$\sigma$ of the mean value were selected. In the bottom left panel of Figure \ref{fig:2} the mean and the dispersion of RV are represented by a black point and error bars. Furthermore, a color-magnitude diagram (CMD) in $G_{\mathrm{bp}}$ - $G_{\mathrm{rp}}$ versus $G_{\mathrm{bp}}$ was made to confirm that all our targets lie along the Red Giant Branch (RGB) or Asymptotic Giant Branch (AGB) of the cluster (colored dots in the bottom left panel of Figure \ref{fig:2}). Finally, only stars with signal-to-noise ratio (S/N) > 50 were chosen to ensure measurement reliability. Although a threshold of 70 is often recommended, previous works (e.g., \citealp{2020ApJ...895...88N}; \citealp{2023A&A...680A..79M}) have shown that reliable results can also be obtained with lower values, especially with {\ttfamily BACCHUS}. Finally, 27 stars satisfied all these selection criteria and were confirmed as our final members. Table \ref{A1} lists the basic parameters of the members stars, i.e., the APOGEE ID, the J2000 coordinates (RA and Dec in degrees), RVs, 2MASS, and \textit{Gaia} DR3 magnitudes.
\section{Methodology}

\subsection{Atmospheric parameters} \label{3.1}

In order to compare different methodologies, we adopted atmospheric stellar parameters obtained through both spectroscopy (with {\ttfamily ASPCAP}) and photometry (employing 2MASS and \textit{Gaia}).

\begin{figure}
\centering
\includegraphics[width=0.49\textwidth, height=0.42\textheight]{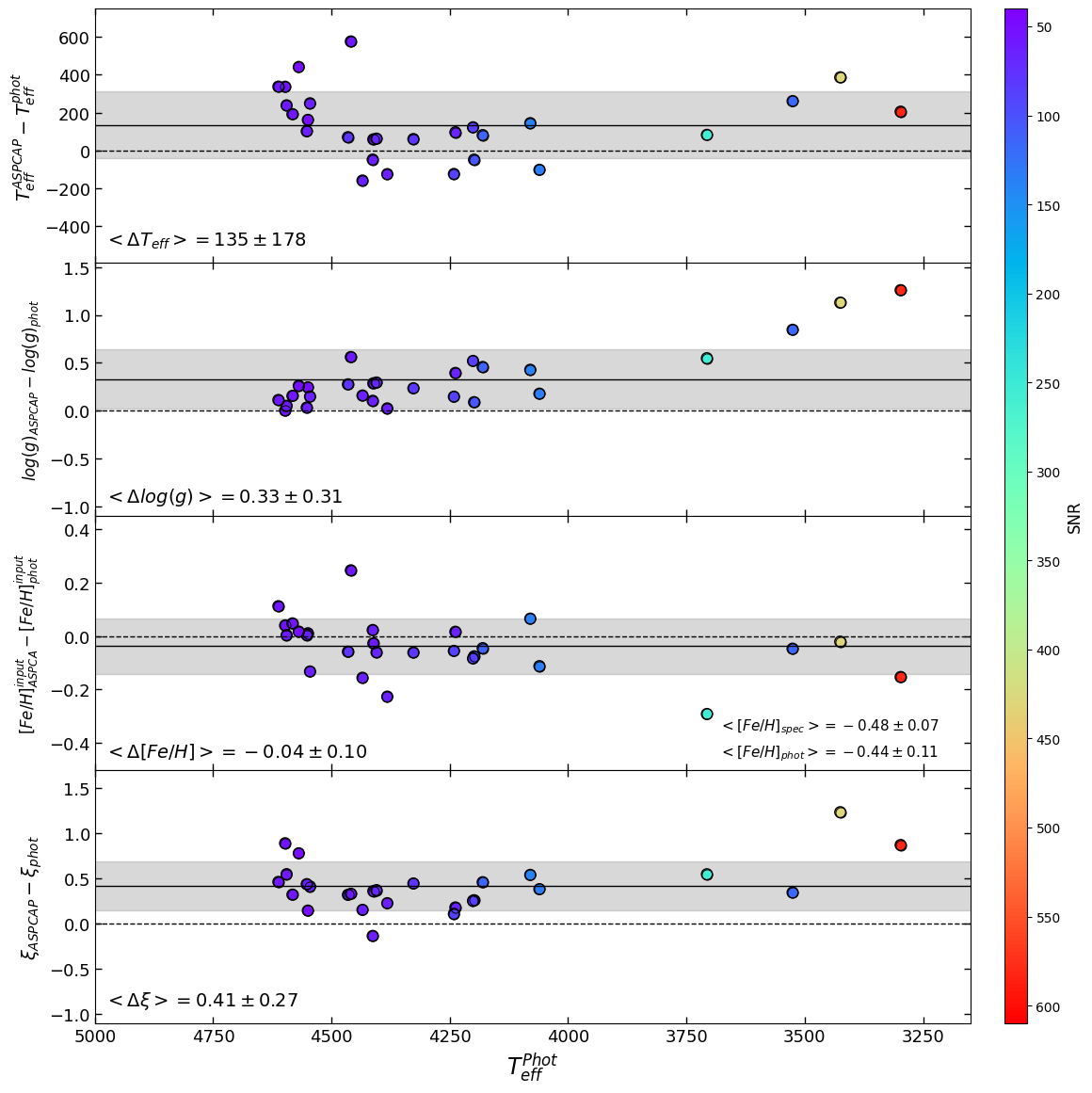}
\caption{Differences in atmospheric parameters, as well as the differences in [Fe/H] resulting from two BACCHUS runs, each adopting different values of $\rm T_{\rm eff}$ and log(g) derived from spectroscopic and photometric parameters. The vertical axis is the difference of each atmospheric parameter and [Fe/H], while the horizontal axis is the $\rm T_{\rm eff}^{\rm phot.}$. Points are color-coded according to S/N. The mean and standard deviation of the differences is represented in each panel by a black line and a gray shadow. These values for each parameter difference are specified at the lower left part of each panel.}
\label{fig:3}
\end{figure}
\subsubsection{Spectroscopic stellar parameters}
We adopted the uncalibrated {\ttfamily ASPCAP} atmospheric parameters. {\ttfamily ASPCAP} uses {\ttfamily MARCS} model atmospheres grid (\citealp{2008A&A...486..951G}) and the {\ttfamily FERRE} code (\citealp{2006ApJ...636..804A}) to perform a $\chi^{2}$ minimization between observed APOGEE-2 spectra and a grid of synthetic spectra to then determine the best-fitting set of parameters and abundances for each spectrum. For further details on {\ttfamily ASPCAP} atmospheric stellar parameters and chemical abundances,  we refer the interested reader to \citet{2018AJ....156..125H} and \citet{2018AJ....156..126J, 2020AJ....160..120J}.

However, it has been observed that {\ttfamily ASPCAP} does not optimally fit the spectra of stars with extreme abundance patterns, particularly 2P stars in GCs. Furthermore, comparison between GC stars and optical spectra references has shown that {\ttfamily ASPCAP} tends to overestimate the effective temperature ($\rm T_{\rm eff}$) and surface gravity (log(g)) for 2P stars. In contrast, for 1P stars, the dispersion in $\rm T_{\rm eff}$ is comparable to that observed in stars with disk-like abundance patterns (\citealp{2018AJ....156..126J, 2020AJ....160..120J}). In addition, \citet{2021A&A...652A.157G, G2025} have observed a trend between the [N/Fe] of 2P stars with [Fe/H] (and $\rm T_{\rm eff}$) compared to 1P. These systematic inaccuracies in atmospheric parameters of 2P stars may introduce significant uncertainties in the chemical abundances derived by APOGEE for GC members. 

To mitigate this limitation, we have also adopted atmospheric parameters derived from photometry as an independent reference to cross validate these results with {\ttfamily ASPCAP}-derived values.

\subsubsection{Photometric stellar parameters}

The atmospheric parameters of our targets were also obtained using photometry through an iterative procedure based on the \textit{Gaia} (G, G$_{\mathrm{BP}}$ and G$_{\mathrm{RP}}$) and 2MASS (J, H, and K$\rm _s$) data. During this procedure, the T$_{\rm eff}$ and log(g) of all RGB stars (including our targets) were obtained, and at the same time, the CMDs were corrected for differential reddening. First, we fit a {\ttfamily PARSEC} isochrone (\citealp{2012MNRAS.427..127B}) to the RGB, assuming an age of 12.0 Gyr (the typical age of a BGC), as shown in Fig. \ref{fig:2}. We accounted for reddening by applying the \citet{1989ApJ...345..245C} reddening law to the isochrone. The visual absorption, A$\rm _V$, the R$\rm _V$ parameter, the intrinsic distance modulus, (m-M)$\rm _0$, and the global metallicity, [M/H], were determined by simultaneously fitting the RGB, the RC, and the RGB tip in the $\rm K_{\mathrm{S}}$ versus G$_{\mathrm{BP}}-K_{\mathrm{S}}$, G versus G$_{\mathrm{BP}}-$G$_{\mathrm{RP}}$, G$_{\mathrm{BP}}$ versus G$_{\mathrm{BP}}-$G$_{\mathrm{RP}}$, and $\rm K_{\mathrm{S}}$ versus $\rm J-K_{\mathrm{S}}$ CMDs. For this purpose only member stars obtained from \textit{Gaia} proper motions were used. T$_{\rm eff}$ and log(g) of each RGB star were then determined as those corresponding to the point on the isochrone where the K magnitude of the star projected along the reddening line intersects the isochrone RGB. We used the K$\rm _s$ magnitude because it is the least affected by reddening and, consequently, by differential reddening. Having the temperature, we obtained the intrinsic G$_{\mathrm{BP}}-K_{\mathrm{S}}$ color of each star from the color-temperature relation of the RGB part of the isochrone and, by subtracting the mean reddening obtained from the isochrone fitting, also the differential reddening at the position of each star. Finally, for each star we selected the four closest neighbors (five stars in total) and corrected its G, G$_{\mathrm{BP}}$, G$_{\mathrm{RP}}$, J, H, and K$\rm _s$, magnitudes using the mean differential reddening of the five stars. We used the G$_{\mathrm{BP}}-K$ color because it is the most sensitive to any reddening variation. This procedure was iterated until any improvement in the CMDs was negligible. We underline the fact that for highly reddened objects such as NGC 6304, the interstellar absorption correction depends on the spectral energy distribution of the star, that is, on its temperature. For this reason, we applied a temperature-dependent absorption correction to the isochrone. Without this, it is not possible to obtain a proper fit to the RGB, especially its upper, cooler part.

We obtained A$_{\rm V}$=1.35, R$_{\rm V}$=2.55, (m-M)$_{\rm 0}$=14.19, and [M/H]=-0.30 dex, a value higher than the iron content of the cluster ([Fe/H]=-0.45). This is not surprising since GCs are usually $\alpha$-enhanced. These extinction parameters correspond to E(B–V)=0.53, in excellent agreement with \citet{1996AJ....112.1487H} value of E(B–V)=0.54. The distance modulus correspond to a distance of 6900 pc.
The microturbulence ($\rm \xi_{\rm t}$) was determined using the empirical equation given in \cite{2016A&A...585A..75D}:
\begin{equation}
\rm \xi_{\rm t} = 0.998 + 3.16 \cdot 10^{-4} X - 0.253 \cdot Y - 2.86 \cdot 10^{-4} X Y +
0.165 \cdot Y^{2}
\end{equation}

where X=T$_{\rm eff}$-5500 K$^{\circ}$ and Y=log(g)-4.0. As \cite{2016A&A...585A..75D} indicate, this equation is consistent with the microturbulence values computed from 3D models. We refer the reader to that paper for detailed information. 

Figure \ref{fig:3} presents the comparison between the spectroscopic and photometric parameters, as well as the differences in [Fe/H] resulting from two {\ttfamily BACCHUS} runs, each adopting different values of $\rm T_{\rm eff}$ and log(g) derived from spectroscopic and photometric parameters. For $\Delta$ $\rm T_{\rm eff}$, we found absolute discrepancies ranging from $\sim$ 48 $\rm K$ to 580 $\rm K$. In the case of $\Delta$ log(g), the absolute differences range from 0.007 to 1.25 observing a systematic difference for the coolest stars. Regarding metallicity, the absolute differences span from 0.004 to 0.29 dex. The mean metallicity difference is <$\Delta$ [Fe/H]> = -0.04 $\pm$ 0.10, suggesting that there are no significant systematic variations. Finally, for $\Delta$ $\rm \xi_{\rm t}$ we found absolute discrepancies on the order of 0.1 - 1.2 km$s^{-1}$. In general, the greatest discrepancies were observed in stars with lower S/N. However, these variations generally do not significantly affect the abundance ratios derived. The stellar spectroscopic and photometric parameters are listed in Table \ref{B2}.

\subsection{Chemical abundances}

For APOGEE spectra, {\ttfamily ASPCAP} determines the chemical abundances of multiple elements for the entire sample. In DR17, {\ttfamily ASPCAP} provides abundances for 17 chemical elements. However, given the peculiar abundances of 2P stars in GCs and the systematic discrepancies in atmospheric parameters and abundances observed by \citet{2018AJ....156..126J, 2020AJ....160..120J} and \citet{2021A&A...652A.157G, G2025}, we re-derive the chemical abundances for NGC 6304 using an independent method in order to cross validate our results with the {\ttfamily ASPCAP} values and provide a consistent chemical analysis.

\subsubsection{The BACCHUS code}

Chemical abundances were derived with the Brussels Automatic Code for Characterizing High-accUracy Spectra ({\ttfamily BACCHUS}; \citealp{2016ascl.soft05004M}), which is based on the {\ttfamily Turbospectrum} radiative transfer code (\citealp{1998A&A...330.1109A}; \citealp{2012ascl.soft05004P}) and the Model Atmospheres with a Radiative and Convective Scheme (\href{http://marcs.astro.uu.se/}{{\ttfamily MARCS}}; \citealp{2008A&A...486..951G}). The abundance of each element is derived using the same procedure described in \citet{2021A&A...652A.158R}, \citet{2023MNRAS.526.6274G} and \citet{2023A&A...673A.123B}, and is summarized as: (i) A synthetic spectra is generated using the full set of atomic and molecular line lists described in \citet{2021AJ....161..254S}. The line list version used is the latest internal DR17, {\ttfamily linelist.20170418}; (ii) From the synthetic spectra, the code finds the local continuum level by a linear fit; (iii) Then, cosmic rays and telluric lines are removed, and the local S/N is estimated; (iv) Subsequently, a series of flux points that contribute to a given absorption line is selected; (vi) Finally, the abundances are derived by comparing the observed spectrum with a set of convolved synthetic spectra characterized by different abundances.

{\ttfamily BACCHUS} includes four different methods that compare the observed spectrum with a set of synthetic spectra of different abundances. In addition, each method has its own quality flags that indicate whether the fit is good (flag = 1) or if the fit is considered problematic (flag $\neq$ 1). These methods are: \\
(1) \textit{Core line intensity comparison (int)} - is based on matching the intensity around the line center between the synthetic and observed spectra. Suitable for blended lines, but also sensitive to adverse effects. \\
(2) \textit{line-profile fitting (syn)} - is the difference between the synthetic line-profile and the observed line-profile. The accuracy of this method depends on the quality of the fit.\\
(3) \textit{equivalent-width comparison (eqw)} - is similar to the syn method and can be affected by badly fit blending lines.\\
(4) \textit{$\chi^{2}$ estimate (chi2)} - determine the abundances by minimizing the squared differences between the synthetic and observed spectra. This approach gives more weight to the core of the line, making it less sensitive to blends in the wings compared to the syn and eqw methods, which apply a flatter weighting over the line.\\
We only considered lines that provided the best fit for the individual abundances. For this reason, we adopted the $\chi^{2}$ method for our abundance determination, as it offers a good balance between strengths and weaknesses for our purpose and ensures consistency with other CAPOS studies. The method is particularly robust when combined with a visual inspection, during which we systematically flagged poorly fit blends, lines affected by nearby contaminating features, or obvious continuum issues. We note, however, that the $\chi^{2}$ method is sensitive to errors in continuum placement, and its results can depend on the S/N of the spectra. Data from other methods were also preserved for comparison and cross-checks.

\subsubsection{Abundance calculations}
\begin{figure}
\includegraphics[width=0.5\textwidth, height=0.66\textheight]{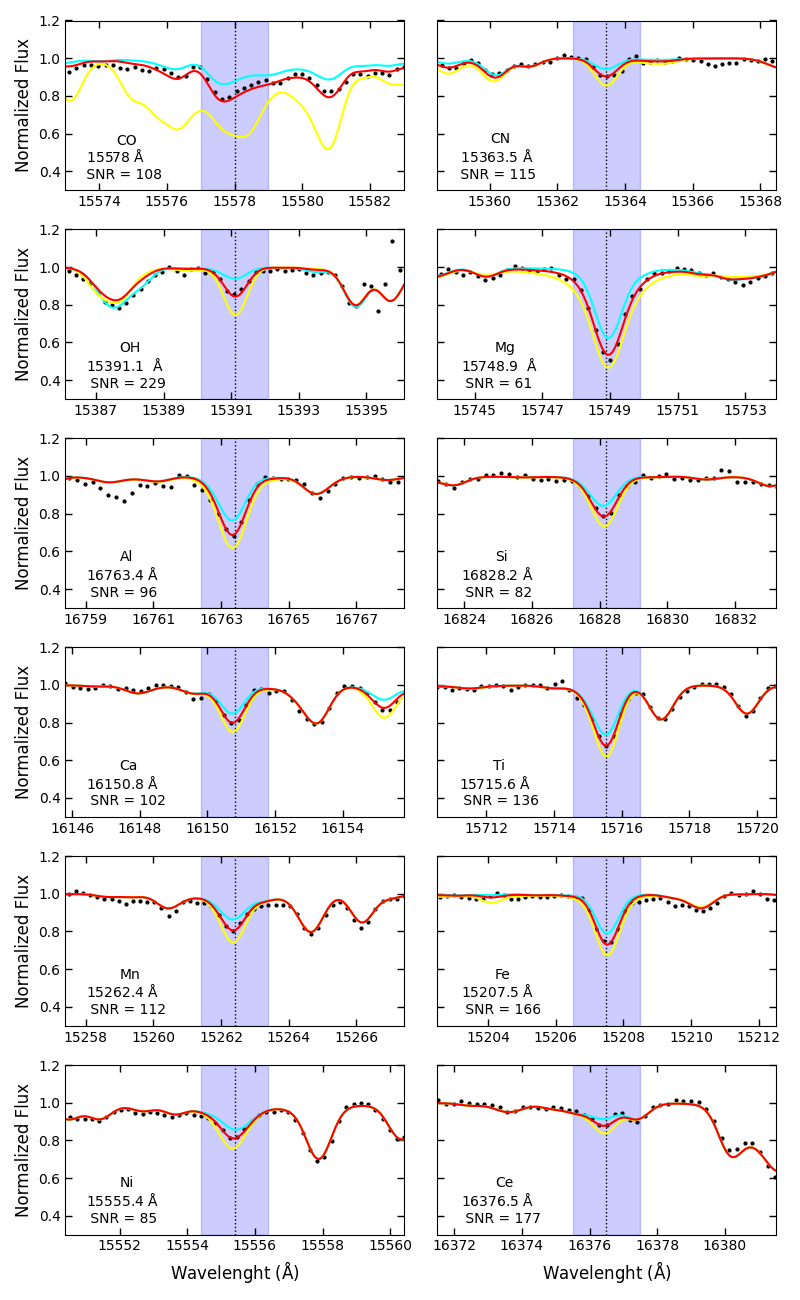}
\caption{Quality of the model fit obtained with {\ttfamily BACCHUS} around the molecular lines ($^{12}$C$^{16}$O, $^{12}$C$^{14}$N, and $^{16}$OH) and atomic lines (Mg, Al, Si, Ca, Ti, Mn, Fe, Ni, and Ce) for the star 2M17143117-2930149 (S/N = 88). The dotted black line represents the observed spectrum, while the solid colored lines correspond to synthetic spectra with abundances varying by $\pm$ 0.3 dex, except for Mg, Al, Si, Ti, and Fe, whose abundances vary by $\pm$ 0.5 dex. The best-fitting synthetic spectrum is shown in red. Each panel is centered on the selected lines with the dashed black lines indicating the positions of the air wavelength lines.}
\label{fig:4}
\end{figure}
The first step using {\ttfamily BACCHUS} consisted of fitting the abundances of C, N and O, with the aim of properly measuring the corresponding molecular features, as these mainly cause the blends in the lines of other species. We first derive the oxygen abundances from the hydroxide molecular lines ($^{16}OH$). With this abundance determined, we proceeded to derive the carbon abundances from the carbon monoxide molecular lines ($^{12}C^{16}O$). Finally, nitrogen abundances were obtained from cyanogen molecular lines ($^{12}C^{14}N$). The C, N, and O abundances were derived iteratively in order to remove any dependence on the OH, CO, and CN lines. Subsequently, we determine the [Fe/H] for each star from selected Fe I lines, using the atmospheric parameters described in Section \ref{3.1} and, as initial guess, the mean [Fe/H] value of {\ttfamily ASPCAP}. Once the [Fe/H] and atmospheric parameters were fixed (see Table \ref{B2}), we proceeded to calculate the abundances of the other elements. In total, we derived 17 chemical species for NGC 6304: C, N, O, Na, Mg, Al, Si, S, K, Ca, Ti, V, Cr, Mn, Fe, Ni and Ce. Figure \ref{fig:4} illustrates the quality of the model fit around selected atomic and molecular lines for the star 2M17143117-2930149, which has a relatively low S/N. The observed spectrum (dotted black line) is compared with synthetic spectra (colored lines), each differing by $\pm$ 0.3 dex in abundance. The best-fitting model is indicated by the red line.

The next step consisted of a visual inspection of the spectrum synthesis output of each element in every star, where we rejected any suspicious fits that had passed the intrinsic {\ttfamily BACCHUS} quality flag. During this inspection, we eliminated lines for which the abundance values derived from the four different methods showed significant discrepancies. After completing the quality selection of lines for each element, we proceeded to calculate the abundance ratio of the elements for each star by averaging the abundances from the selected lines. The adopted solar abundances are taken from \citet{2007SSRv..130..105G}. The resulting chemical abundances that we measure with {\ttfamily BACCHUS} are reported in Tables \ref{tab:C1} and \ref{tab:C2}, while those {\ttfamily ASPCAP} abundances are reported in Table \ref{tab:C3}.
\subsection{Errors} \label{3.3}

\begin{figure}
\includegraphics[width=0.5\textwidth, height=0.17\textheight]{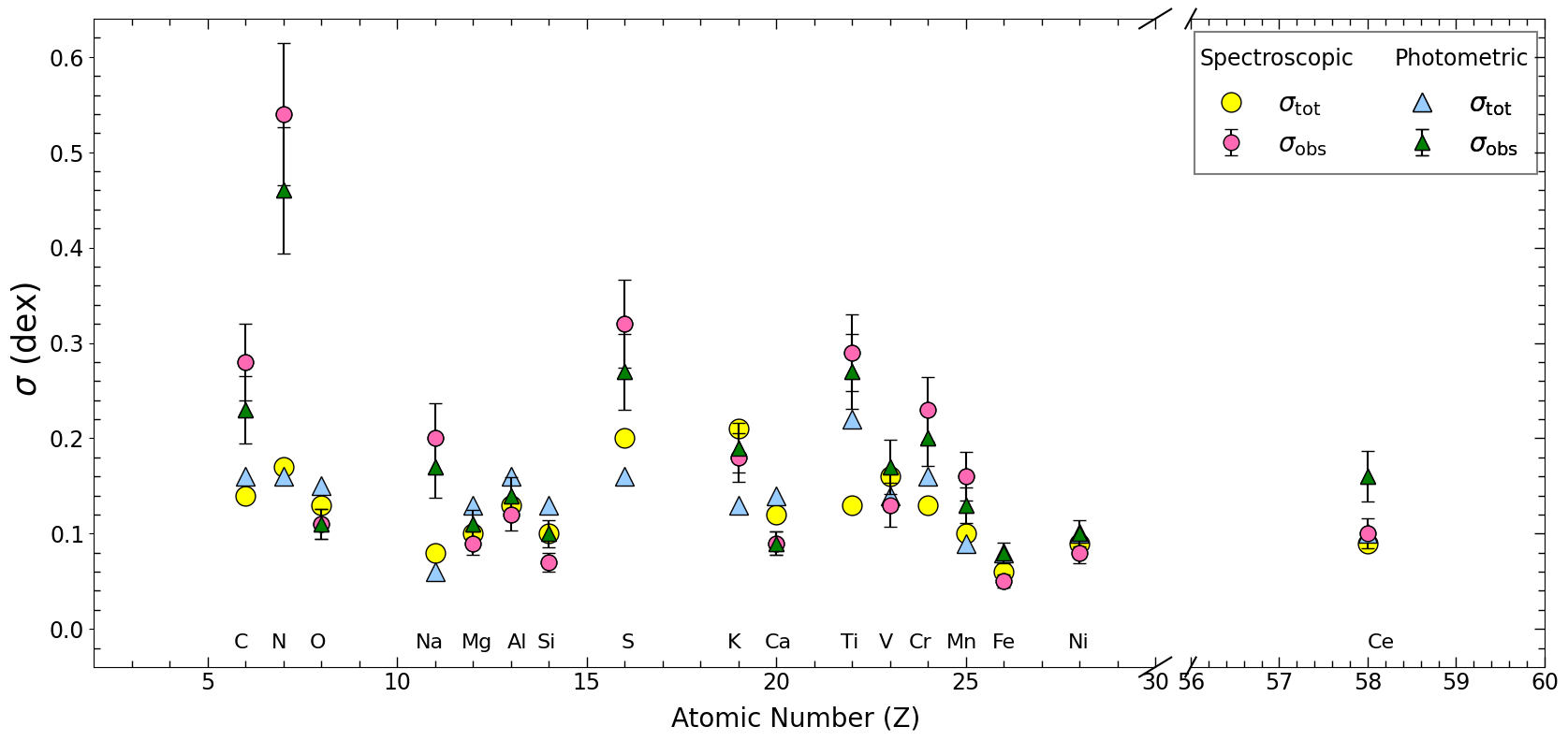}
\caption{Comparison between $\sigma_{\rm tot}$ and $\sigma_{\rm obs}$ as a function of the atomic number (Z). Total error of our measurements and the observed error derived from {\ttfamily BACCHUS} using spectroscopic stellar parameters are represented by yellow and pink circles, respectively, while those obtained using photometric parameters are shown as blue and green triangles. Error bars indicate the error of the mean. Element abbreviations are shown for clarity.}
\label{fig:5}
\end{figure}

The uncertainties associated with our stellar parameters were estimated by recalculating the elemental abundances while varying each parameter individually by the following values: $\Delta \rm T_{\rm eff}$ = $\pm$ 115 K, $\Delta$log(g) = $\pm$ 0.25 dex, $\Delta$[Fe/H] = $\pm$ 0.03 dex, and $\Delta \xi_{\rm t}$ = $\pm$ 0.35 [Km$\rm{s}^{-1}$]. These variations were selected by comparing the spectroscopic and photometric parameters, considering a 1.5$\sigma$ range from the mean value. We then selected a star (2M17141615-2925204) as representative of the entire sample in terms of its parameters and systematically varied its parameters one at a time to recalculate the elemental abundances, generating eight different abundance values per element. Next, we calculate a standard deviation for each element that includes the original abundance and the two abundances obtained after varying the parameter, as follows:
\begin{table}[H]
\centering
\begin{tabular}{c}
$\displaystyle
\begin{aligned}
\sigma_{\rm T_{\rm eff}} &= \sigma\left([X/\rm{Fe}], [X/\rm{Fe}]^{T_{\rm eff}}_{+115}, [X/\rm{Fe}]^{T_{\rm eff}}_{-115}\right) \\

\sigma_{\log g} &= \sigma\left([X/\rm{Fe}], [X/\rm{Fe}]^{\log g}_{+0.25}, [X/\rm{Fe}]^{\log g}_{-0.25}\right) \\

\sigma_{[\rm{Fe}/\rm{H}]} &= \sigma\left([X/\rm{Fe}], [X/\rm{Fe}]^{[\rm{Fe}/\rm{H}]}_{+0.25}, [X/\rm{Fe}]^{[\rm{Fe}/\rm{H}]}_{-0.25}\right) \\

\sigma_{\xi_{\rm t}} &= \sigma\left([X/\rm{Fe}], [X/\rm{Fe}]^{\xi_{\rm t}}_{+0.35}, [X/\rm{Fe}]^{\xi_{\rm t}}_{-0.35}\right).
\end{aligned}
$
\end{tabular}
\end{table}
Additionally, we estimated the error due to the noise in the spectra ($\sigma_{\rm{S/N}}$), by dividing the rms scatter by the square root of the number of lines used for a given element and star. Finally, the total error ($\sigma_{\rm{tot}}$), in our abundance measurements, is given by the relation
\begin{table}[H]
\centering
\begin{tabular}{c}
$\displaystyle
 \sigma_{\rm{tot}} = \sqrt{\sigma^2_{\rm{T_{eff}}} + \sigma^2_{\rm{\log(g)}} + \sigma^2_{\rm{\xi_{t}}} + \sigma^2_{\rm{[Fe/H]}} + \sigma^2_{\rm{S/N}}}  
$ .
\end{tabular}
\end{table}
The errors for each [X/Fe] ratio, resulting from uncertainties in both spectroscopic and photometric stellar parameters, as well as $\sigma_{\rm S/N}$, and the uncertainty provided by {\ttfamily ASPCAP} ($\sigma_{\rm ASPCAP}$) are listed in Table \ref{tab:D1}. However, special caution should be taken with the uncertainties for Na. In some cases, the standard deviation associated with the parameter variation could only be calculated from two values instead of three because of the absence of a calculated abundance. Additionally, only two Na I lines are present in the APOGEE spectral range, of which only one was considered moderately measurable. Thus, these factors contribute to a possible underestimation of the true uncertainty associated with Na. In Figure \ref{fig:5}, we compare $\sigma_{\rm tot}$ with the observed dispersion ($\sigma_{\rm obs}$) obtained using spectroscopic and photometric stellar parameters as function of the atomic number, to further assess the impact of the uncertainties of these parameters on our abundance determination. The error bars represent the error of the observed dispersion ($rms/\sqrt{2N}$, where N is the number of stars). In this comparison, we observe that, in general, photometric parameters exhibit a smaller dispersion than spectroscopic ones. In addition, we also see a significant difference between $\sigma_{\rm tot}$ and $\sigma_{\rm obs}$ for C, N, and Na, suggesting the presence of an intrinsic spread in these elements within the cluster. In contrast, Mg and Al show relatively small differences, which is consistent with the absence of intrinsic spread of these elements in metal-rich GCs (\citealp{2017A&A...601A.112P}; \citealp{2020MNRAS.492.1641M}). Other elements such as S, Ti, and Cr also exhibit a $\sigma_{\rm obs}$ larger than $\sigma_{\rm tot}$. However, only a limited number of lines are considered reliable for these elements, which limit the accuracy of the abundances and may account for the difference observed. Similarly, we can see that K and Ce show discrepancies when photometric parameters are used but not when spectroscopic parameters are adopted. These inconsistencies can be attributed to the sensitivity of the lines to the input stellar parameters adopted. On the other hand, we see good agreement for the other elements.

The internal errors of {\ttfamily ASPCAP} are described in \citet{2016AJ....151..144G} and are estimated from the inverse of the curvature matrix of the elements. This matrix depends on the flux error and the partial derivatives of the synthetic spectra with respect to the different parameters/abundances considered in the minimization algorithm $\chi^{2} $. However, the abundance uncertainties derived in star clusters are generally larger than the internal {\ttfamily ASPCAP} uncertainties (\citealp{2015AJ....150..148H}). In subsequent data releases (DR13, DR14, and DR16), these uncertainties were adjusted by repeat observations of a set of targets (\citealp{2018AJ....156..125H};  \citealp{2020AJ....160..120J}). The difference between the abundances of different {\ttfamily ASPCAP} estimates provides information on the uncertainties. Therefore, these repeat observations were used to estimate the uncertainty as a function of $T_{\mathrm{eff}}$, [M/H] and S/N. For DR17, a general description of the results and uncertainties  will be provide by Holtzman et al. (2022, in preparation).

Consequently, {\ttfamily ASPCAP} uncertainties are generally smaller than those derived from our {\ttfamily BACCHUS} error analysis using both sets of atmospheric parameters.

\section{Results}

In the following section, the chemical abundances of iron, $\alpha$, and iron-peak elements are represented.

\subsection{Iron}

Taking into account the {\ttfamily ASPCAP} value together with {\ttfamily BACCHUS} values derived from spectroscopic and photometric stellar parameters, we obtained a mean metallicity for NGC 6304 of:
\begin{table}[H]
\centering
\begin{tabular}{c}
$\displaystyle
\begin{aligned}
[\rm Fe/H]_{\rm ASPCAP} = -0.49\pm0.05\\
[\rm Fe/H]_{\rm spec} = -0.45\pm0.05\\
[\rm Fe/H]_{\rm phot} = -0.45\pm0.08\\
\end{aligned}
$
\end{tabular}
\end{table}
with the error being the standard deviation of the mean. By comparing these values with our uncertainties in each method, we did not find any intrinsic spread for [Fe/H]. Using spectroscopic parameters, we see $\sigma_{\rm tot}$ = 0.06, which is in good agreement with $\sigma_{\rm obs}$ = 0.05. While, for the photometric parameters, we see $\sigma_{\rm tot}$ = 0.08, which is in very good agreement with $\sigma_{\rm obs}$ = 0.08. 

Other studies based on APOGEE DR17 data for this cluster include \citet{2024MNRAS.528.1393S}, who derived a value of [Fe/H] = -0.48 from a sample of 34 stars, and \citet{G2025}, who reported a value of [Fe/H] = -0.49 using {\ttfamily ASPCAP} and only stars with S/N > 70. To our knowledge, no other high-resolution spectroscopic studies of NGC 6304 have been published. However, there are some photometric studies: \citet{2005MNRAS.361..272V}, using near-IR photometry, derived a lower metallicity of [Fe/H] = –0.70, \citet{2020ApJ...891...37O} obtained a value of [Fe/H] = –0.48 from the isochrone fitting based on HST photometry, and \cite{2000A&A...357..495O} using the B and V bands, reported a metallicity of [Fe/H] = –0.57. In summary, our mean metallicity determination for NGC 6304 agrees well with the various APOGEE spectroscopic estimates and is broadly consistent with previous photometric determinations. Finally, \cite{1996AJ....112.1487H} list a value of [Fe/H] = –0.45, which is in excellent agreement with our estimation.

\subsection{$\alpha$ elements}

The $\alpha$ elements measured in this study are O, Mg, Si, S, Ca, and Ti (see Tables \ref{tab:C1}, \ref{tab:C2}, and \ref{tab:C3}). Considering Mg, Si, and Ca as our most reliable elements (see above), we obtain a mean $\alpha$ abundance of:
\begin{table}[H]
\centering
\begin{tabular}{c}
$\displaystyle
\begin{aligned}
[\rm \alpha/Fe]_{\rm ASPCAP} = 0.21\pm0.05\\
[\rm \alpha/Fe]_{\rm spec} = 0.24\pm0.07\\
[\rm \alpha/Fe]_{\rm phot} = 0.23\pm0.08\\
\end{aligned}
$
\end{tabular}
\end{table}
The error is the standard deviation of the mean. The abundance of O was not considered because it is affected by MP (see Section \ref{5}). Similarly, S and Ti were also excluded due to their lower reliability (see Section \ref{Appendix E}). In contrast, the elements Mg, Si, and Ca show good reliability and also we did not find significant dispersions. 

Individually, {\ttfamily ASPCAP} provides lower values compared to our {\ttfamily BACCHUS} estimates for these elements. For instance, the mean value of Mg reported by {\ttfamily ASPCAP} ([Mg/Fe] = 0.23) is 0.06 dex lower than our {\ttfamily BACCHUS} estimations using both parameters ([Mg/Fe]$_{\rm spec, \rm phot}$ = 0.29). A slightly larger discrepancy is found for Si, where the mean abundance derived by {\ttfamily ASPCAP} ([Si/Fe] = 0.21) is significantly lower than our values ([Si/Fe]$_{\rm spec}$ = 0.29 and [Si/Fe]$_{\rm phot}$ = 0.31), reaching a difference of up to 0.1 dex. This discrepancy may suggest the presence of systematic effects between the two techniques. Similar results are discussed in \citet{G2025} for other CAPOS clusters reduced with {\ttfamily BACCHUS}. An exception is found for Ca, where our results ([Ca/Fe]$_{\rm spec}$ = 0.14; [Ca/Fe]$_{\rm phot}$ = 0.10) are lower than those reported by {\ttfamily ASPCAP} ([Ca/Fe] = 0.18). Other APOGEE $\alpha$ abundances reported for NGC 6304 are [$\alpha$/Fe] = 0.20$\pm0.04$ and [$\alpha$/Fe] = 0.20$\pm0.03$, respectively (\citealp{G2025}; \citealp{2024MNRAS.528.1393S}). In general, our results are consistent with these previous estimates.

\begin{figure}
\includegraphics[width=0.48\textwidth, height=0.32\textheight]{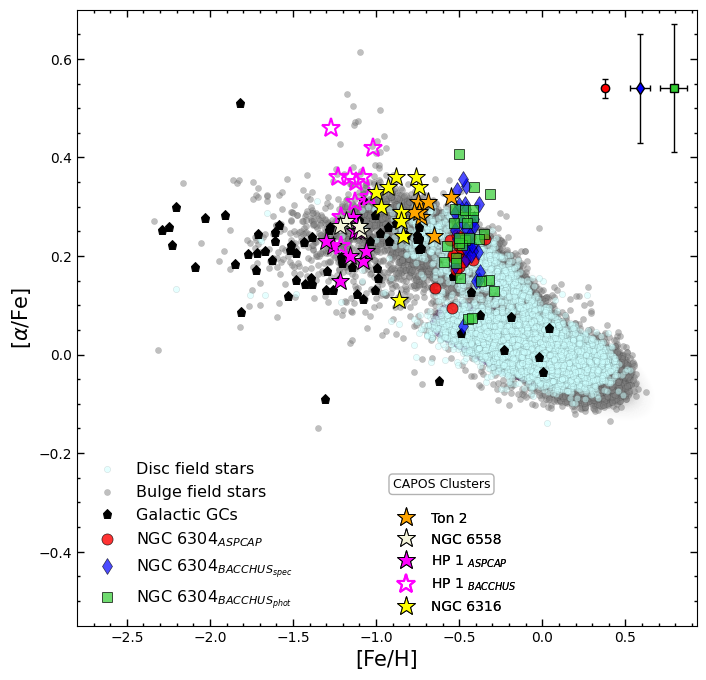}
\caption{[$\alpha$/Fe] vs. [Fe/H]. Filled red circles, blue diamonds and green squares are our data for NGC 6304, considering the three methods used. Stars represent different CAPOS CGs, orange: Ton 2 (\citealp{2022A&A...658A.116F}), beige: NGC 6558 (\citealp{2023MNRAS.526.6274G}), filled magenta: HP 1 from ASPCAP, empty magenta: HP 1 from BACCHUS (\citealp{2025A&A...696A.154H}), yellow: NGC 6316 (\citealp{2025A&A...701A.159F}). Filled black pentagon: Galactic GCs (\citealp{2024MNRAS.528.1393S}). Filled cyan circles: disk field stars (\citealp{2015ApJ...808..132H}). Filled gray circles: bulge field stars (\citealp{2020MNRAS.499.1037R}).}
\label{fig:6}
\end{figure}

Figure \ref{fig:6} shows the $\alpha$ elements considered as a function of [Fe/H]. From this figure, we can see that the $\alpha$ abundances are overabundant with respect to the Sun, as found for almost all GCs, except for a few young GCs associated with the Sagittarius dwarf galaxy and Rup 106 (\citealp{G2025}). In addition, we compare NGC 6304 with other CAPOS clusters: Ton 2 (\citealp{2022A&A...658A.116F}), NGC 6558 (\citealp{2023MNRAS.526.6274G}), HP 1 (\citealp{2025A&A...696A.154H}), and NGC 6316 (\citealp{2025A&A...701A.159F}), which were classified as BGCs by \citet{G2025}, with the exception of Ton 2 which they classified as thick disk, as well as \citet{2020MNRAS.491.3251P}. Moreover, we also compare with other Galactic GCs (\citealp{2024MNRAS.528.1393S}), disk field stars (\citealp{2015ApJ...808..132H}) and bulge field stars (\citealp{2020MNRAS.499.1037R}). The last two are updated to DR17. From this comparison, we observe that NGC 6304 is in good agreement with both bulge field stars and other bulge GCs, suggesting a similar chemical enrichment history.
\begin{figure}
\begin{center}
\includegraphics[width=0.48\textwidth]{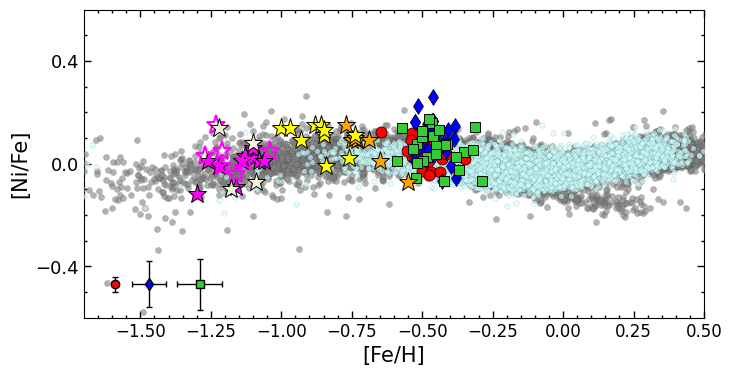}
\caption{[Ni/Fe] vs. [Fe/H]. The symbols and color codes for NGC 6304, CAPOS clusters and field stars are the same as in Fig. \ref{fig:6}.} 
\label{fig:7}
\end{center}
\end{figure}
\subsection{Iron peak}

The iron-peak elements measured in this study are V, Cr, Mn, and Ni (see Tables \ref{tab:C1}, \ref{tab:C2}, and \ref{tab:C3}). 
According to our spectral line quality classification (see Appendix \ref{Appendix E}), the abundance determinations for V, Cr, and Mn have intermediate reliability, mainly due to the limited number and weakness of their available lines. This classification does not imply that the derived abundances are unreliable, but rather that they have larger associated uncertainties compared to those classified as good. For completeness, we include their values in the abundance tables, although our discussion primarily focuses on elements with higher-quality determinations. Consequently, Ni is the only iron-peak element for which we report a good abundances measurement (see Appendix \ref{Appendix E}). The mean Ni abundances measured with {\ttfamily ASPCAP} and {\ttfamily BACCHUS} are:
\begin{table}[H]
\centering
\begin{tabular}{c}
$\displaystyle
\begin{aligned}
[\rm Ni/Fe]_{\rm ASPCAP} = 0.04\pm0.04\\
[\rm Ni/Fe]_{\rm spec} = 0.10\pm0.08\\
[\rm Ni/Fe]_{\rm phot} = 0.06\pm0.07\\
\end{aligned}
$
\end{tabular}
\end{table}
where the error is the standard deviation of the mean. \citet{G2025} measured a value of [Ni/Fe] = 0.07$\pm0.03$, while \citet{2024MNRAS.528.1393S} derived a [Ni/Fe] = 0.04$\pm0.05$. In general, our values are in concordance with the previous results. Figure \ref{fig:7} shows [Ni/Fe] versus [Fe/H], compared to the same CAPOS clusters, bulge field stars, and disk field stars in Figure \ref{fig:6}. From Figure \ref{fig:7}, we can see that the Ni abundance in NGC 6304 agrees well with the bulge field stars and is consistent with the other BGCs.

\section{Multiple population} \label{5}

The strongest evidence for the presence of MP in GCs is the star-to-star variation in the abundances of light elements (C, N, O, Na, Mg, and Al), observed in almost all Galactic GCs. These elements are produced at high temperatures through proton capture reactions associated with hydrogen burning. At temperatures of approximately 15 $\times10^{6}$ K, the CNO cycle is activated, converting C and O into N. At around 30 $\times10^{6}$ K, the NeNa chain becomes active, producing Na at the expense of Ne. Finally, at even higher temperatures ($\sim$ 70 $\times10^{6}$ K), the MgAl chain is activated, leading to the production of Al at the expense of Mg (\citealp{2007A&A...470..179P}).

\begin{figure*}
\begin{center}
\includegraphics[width=0.94\textwidth, height=0.48\textheight]{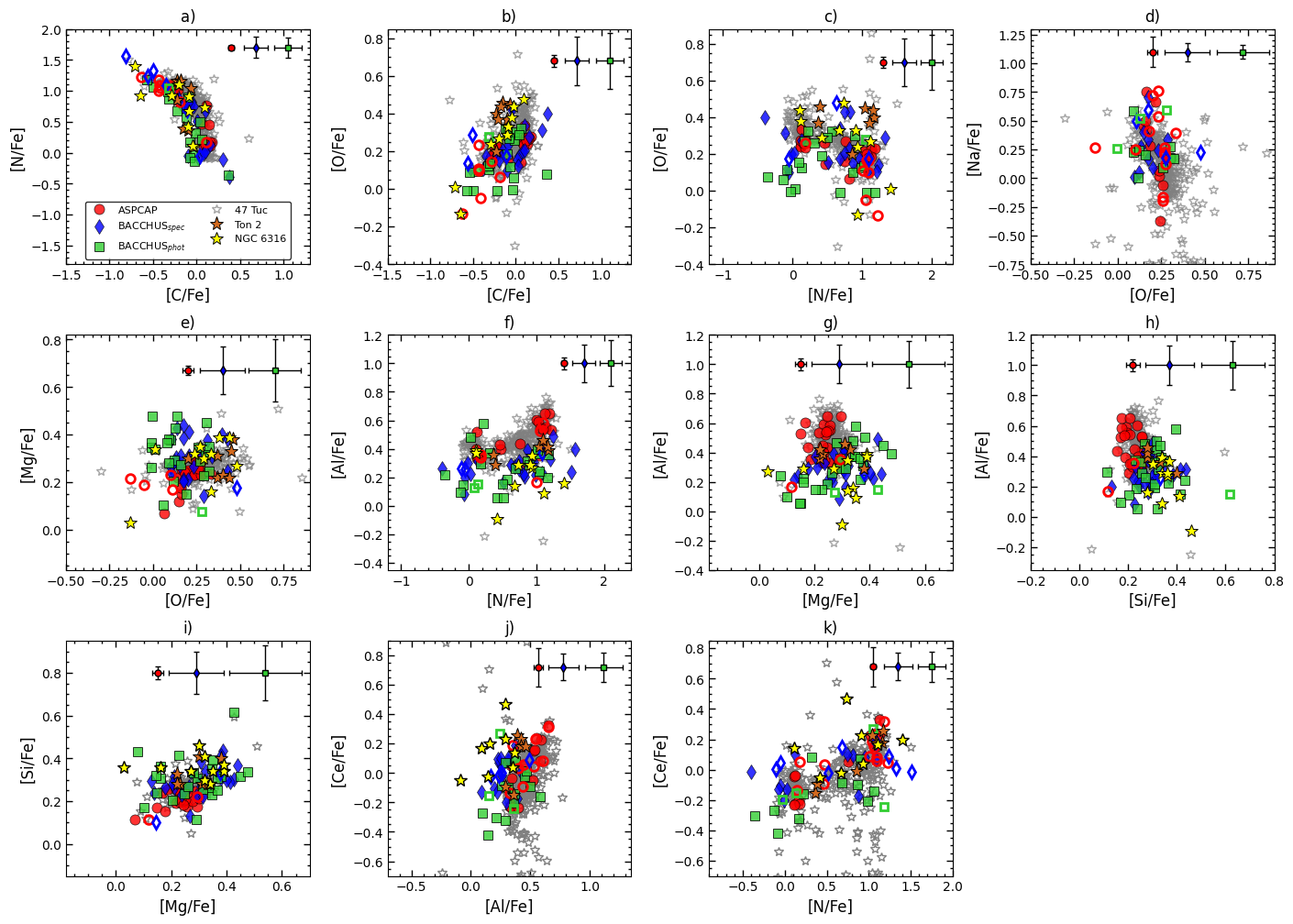}
\caption{Light and s-process element distributions for NGC 6304. Filled red circles, blue diamonds, and green squares represent the common stars in NGC 6304 where all three methods could be performed. The empty red circles, blue diamonds, and green squares denote the remaining stars analyzed with each method individually. For comparison we show other GCs of similar metallicity: 47 Tuc (gray stars; \citealp{2024MNRAS.528.1393S}) and two GCs of the CAPOS sample: Ton 2 (orange stars; \citealp{2022A&A...658A.116F}) and NGC  6316 (yellow stars; \citealp{2025A&A...701A.159F}). The errors associated with each abundance are included in each plot (for {\ttfamily ASPCAP} error see Section \ref{3.3}).} 
\label{fig:8}
\end{center}
\end{figure*}

In Figure \ref{fig:8}, we present the abundance relations of these light elements derived from our analysis, along with Ce, in order to explore possible correlations between Ce-Al and Ce-N. Additionally, we compare NGC 6304 with other GCs of similar metallicity observed with APOGEE:  Ton 2 ([Fe/H] = -0.70; \citealp{2022A&A...658A.116F}) and NGC 6316 ([Fe/H] = -0.87; \citealp{2025A&A...701A.159F}) from CAPOS and 47 Tuc ([Fe/H] = -0.74; \citealp{2024MNRAS.528.1393S}).
\subsection{C-N anticorrelation}

The C and N abundances are influenced by two distinct astrophysical processes. The first happens on the surface of evolved giant stars, where the first dredge-up occurs, bringing nuclear-processed material from the stellar core to the surface. However, since all of our stars are located well above the RGB bump (where the first dredge-up takes place), this process alone is insufficient to explain the C-N anticorrelation observed in GCs (\citealp{2007A&A...470..179P}). This chemical pattern is instead attributed to a second process: the pollution by a previous generation of massive stars, where the 2P of stars was enriched from the processed material of 1P stars (\citealp{2011A&A...533A.120V}; \citealp{2018ARA&A..56...83B}). C-N anticorrelation is observed in nearly all GCs and is a key indicator of the presence of MP.

In panel a) of Figure \ref{fig:8} we see a C-N anticorrelation for NGC 6304 with a strong enrichment in N and a depletion in C. This means that as the N abundance increases, the C abundance decreases, reflecting the classical signature of the CNO-cycle processing. A significant star-to-star spread in N is evident in both our spectroscopic and photometric parameter analysis. Using spectroscopic parameters, we find $\sigma_{\rm obs}$ = 0.54, which is much higher than $\sigma_{\rm tot}$ = 0.17. A similar trend is found with the photometric parameters, where $\sigma_{\rm obs}$ = 0.46 compared to $\sigma_{\rm tot}$ = 0.16. This chemical pattern is consistent with those observed in the other GCs shown in Figure \ref{fig:8}.

\subsection{Na-O anticorrelation}

The Na-O anticorrelation is one of the most important chemical signatures of MP in GCs and has been observed in many Galactic GCs (\citealp{2009A&A...505..117C}; \citealp{2012A&A...544A..12G}). This anticorrelation is caused by the CNO cycle depleting O and by the NeNa-chain enriching Na. Several studies have shown that the extension of the Na-O anticorrelation is related with the mass and metallicity of metal-poor halo GCs (\citealp{2009A&A...505..117C, 2010A&A...516A..55C}, \citealp{2017A&A...601A.112P}). In particular, a clear and extended Na-O anticorrelation is observed in massive and metal-poor GCs. However, this is not the case for metal-rich BGCs. There are studies that have observed significant star-to-star variation in Na, but little or no variation in O (\citealp{2017A&A...605A..12M, 2018A&A...620A..96M, 2020MNRAS.492.3742M}; \citealp{2021MNRAS.503.4336M}). This behavior suggests that BGCs follow the same relationship between the mass and metallicity with their Na-O anticorrelation extension as observed for halo and disk GCs, with the caveat that BGCs reach higher metallicities than in these other GC systems, where the metallicity effect becomes more apparent. In panel d) of Figure \ref{fig:8}, we plot for the first time the abundance ratios of Na and O for NGC 6304 with high-resolution spectroscopy. It is important to emphasize that Na abundances are particularly difficult to determine in APOGEE spectra because of very few, weak Na I lines. However, we consider that at least one of the two available Na I lines is moderately measurable, and the results for NGC 6304 are consistent with what is expected from the literature. We find that NGC 6304 follows the same pattern as observed for other high-metallicity BGCs, characterized by a significant spread for Na and minimal variation in O. This trend is observed for both spectroscopic and photometric analyses, as well as for {\ttfamily ASPCAP} results. Using spectroscopic parameters, we obtain for Na a spread of $\sigma_{\rm obs}$ = 0.20 and a total uncertainty of $\sigma_{\rm tot}$ = 0.08, whereas for O we find $\sigma_{\rm obs}$  = 0.11 and $\sigma_{\rm tot}$ = 0.13. Using photometric parameters, we derive $\sigma_{\rm obs}$  = 0.17 and $\sigma_{\rm tot}$ = 0.06 for Na, and $\sigma_{\rm obs}$  = 0.11 and $\sigma_{\rm tot}$ = 0.15 for O. Compared with 47 Tuc, classified as an inner-halo GC by \citet{2020MNRAS.491.3251P}, we see that the Na-O anticorrelation is more pronounced, but we note that 47 Tuc is some 0.25 dex more metal-poor.

\subsection{Mg-Al anticorrelation}

The Mg-Al anticorrelation is not observed in all GCs \citealp{2015AJ....149..153M, 2020MNRAS.492.1641M}). This is because the MgAl chain requires much higher temperatures to operate ($\sim$70 million K), which are only achieved in the cores of low-metallicity polluter stars. Additionally, similar to the Na-O anticorrelation, the extension of the Mg-Al anticorrelation appears to strongly depend on both the mass and the metallicity of the GC. Specifically, metal-poor GCs tend to exhibit a well-developed and curved Mg-Al anticorrelation, while more metal-rich GCs do not show clear signs of this anticorrelation (\citealp{2009A&A...505..117C}, \citealp{2017A&A...601A.112P}).

No evidence of a Mg-Al anticorrelation is observed for NGC 6304, as shown in panel g) of Figure \ref{fig:8}. This result is consistent for both the spectroscopic and photometric analyses, as well as for the {\ttfamily ASPCAP} results. The observed spreads in Al and Mg are essentially the same. With spectroscopic parameters we obtain for Al a $\sigma_{\rm obs}$ = 0.12 and a $\sigma_{\rm tot}$ = 0.13, while for Mg we find $\sigma_{\rm obs}$ = 0.09 and $\sigma_{\rm tot}$ = 0.10. Using photometric parameters, we observed a $\sigma_{\rm obs}$ = 0.14 and $\sigma_{\rm tot}$ = 0.16 for Al, and $\sigma_{\rm obs}$ = 0.11 and $\sigma_{\rm tot}$ = 0.13 for Mg. The other GCs shown in panel g) exhibit the same behavior as NGC 6304. In contrast, NGC 6316 shows a significant star-to-star variation in both Al and Mg. This result is consistent with that established by \citet{2009A&A...505..139C}, where the Mg-Al anticorrelation is typically observed in clusters that are more massive (M>10$^{5}$ M$_{\odot}$), metal-poor ([Fe/H] < -1.5), or both. According to \href{https://people.smp.uq.edu.au/HolgerBaumgardt/globular/}{Baumgardt database}, NGC 6304 has a mass that meets the threshold for a massive GC, but is very metal-rich. A similar case is 47 Tuc, with a higher mass of 8.53$\pm$0.05 $\times10^{5}$M$_{\odot}$, but also classified as a metal-rich GC. Ton 2 is also metal-rich and has a relatively low mass of 4.31$\pm$0.91$\times 10^{4}$M$_{\odot}$. The case of NGC 6316 is different, although it is not strictly metal-poor, its metallicity of -0.87 (\citealp{2025A&A...701A.159F}) places it closer to the intermediate-metallicity regime and it has a sufficiently high mass of 3.47$\pm$0.44$\times 10^{5}$M$_{\odot}$ to be considered a massive cluster. These results support the trend that metal-rich GCs do not show a Mg-Al anticorrelation, suggesting that metal-rich GCs experience different nucleosynthesis pathways from those of metal-poor ones (\citealp{2020MNRAS.492.1641M}).

\subsection{Cerium}

Ce is mainly produced by neutron capture reactions (n-capture), which occurs via two processes: the slow n-capture process (s-process), in which the timescale for n-capture is much longer than the beta-decay lifetime, and the rapid n-capture process (r-process), where the timescale of the n-capture is much shorter than the beta-decay lifetime. The s-process occurs primally in AGB stars, while the r-process is associated with SNe explosions. Ce is produced through a combination of both the r- and s-processes. 

The possible association of Ce with MP emerged from \citet{2021ApJ...918L...9F}, who observed a correlation between Ce, N and Al in NGC 6380 ([Fe/H] = -0.80), which was the first GC in which Ce-N and Ce-Al correlations were detected. Subsequently, \citet{2022A&A...658A.116F} found similar correlations in Ton 2. However, \citet{2023MNRAS.526.6274G} and \citet{2025A&A...696A.154H} did not find evidence for such correlations in NGC 6558 and HP 1, perhaps subject to the limited number of stars analyzed per cluster. More recent studies have confirmed the presence of Ce-N and Ce-Al correlations in NGC 6569 (\citealp{2025A&A...699A.128B}) and NGC 6316 (\citealp{2025A&A...701A.159F}) but not in Terzan 2 (\citealp{Uribe2025}). In NGC 6304, a clear correlation is seen in panels j) and k) of Figure \ref{fig:8}. This pattern is observed in the {\ttfamily ASPCAP} results and with photometric parameters. However, the correlation is less pronounced in {\ttfamily BACCHUS} when the spectroscopic parameters were considered. Specifically, when using spectroscopic parameters, we find a dispersion in Ce of $\sigma_{\rm obs}$ = 0.10 and $\sigma_{\rm tot}$ = 0.09. However, when using photometric parameters, we derive $\sigma_{\rm obs}$ = 0.16 and $\sigma_{\rm tot}$ = 0.10. Similar correlations of Ce, N, and Al are observed for 47 Tuc.

The observed Ce–N and Ce–Al correlations in several metal-rich BGCs suggest that Ce may serve as an additional tracer of MP. Moreover, these correlations suggest a scenario in which the enrichment of these elements occurred through similar astrophysical processes, possibly related to the contribution of AGB stars. In this context, AGB stars contribute not only for the enrichment of light elements such as N and Al through hot bottom burning, but also for the production of s-process elements such as Ce, supporting the hypothesis that AGB stars played a significant role in the chemical enrichment of these clusters (\citealp{2016ApJ...831L..17V}).
\section{Concluding remarks}

In this article, we present for the first time a detailed chemical analysis of the BGC NGC 6304. We analyzed 27 high-probability RGB members observed in the H band using the APOGEE-2 spectrograph, as part of the CAPOS survey. To analyze the spectra, we used two different sets of stellar parameters: one directly from {\ttfamily ASPCAP} and another derived from photometric data (2MASS and \textit{Gaia}). Using the {\ttfamily BACCHUS} code, we measured abundances for 17 elements, including light (C and N), $\alpha$ (O, Mg, Si, S, Ca and Ti), odd-Z (Al, Na and K), iron-peak (V, Cr, Mn, Fe and Ni), and the s-process element, Ce. For each element, we assessed the quality of the spectral lines and assigned them to one of three categories according to their reliability. We also carried out a detailed error analysis, comparing how the observed error varied in comparison to the total error that we found. Subsequently, we compared the chemical patterns of NGC 6304 with those of other BGCs in the CAPOS survey, as well as other GCs and field stars from the bulge, disk, and halo of the MW. Finally, we performed a study of the MP in NGC 6304, focusing on the typical abundance variations including the N-C, Na-O, and Mg-Al anticorrelations, as well as the Ce correlations with N and Al. The main results of our study are presented below.
\begin{enumerate}

\item The mean metallicity that we derived for NGC 6304 using {\ttfamily BACCHUS}, with spectroscopic and photometric stellar parameters is: $[{\rm Fe/H}]_{\rm spec} = -0.45 \pm 0.05$ and $[{\rm Fe/H}]_{\rm phot} = -0.45 \pm 0.08$, respectively. These values agree well with the result of {\ttfamily ASPCAP} of $[{\rm Fe/H}]_{\rm ASPCAP} = -0.49 \pm 0.05$, and is in concordance with previous studies reported in the literature. Furthermore, we did not find any significant spread in iron content.

\item $\alpha$ elements in NGC 6304 are overabundant with respect to the Sun. Considering Mg, Si, and Ca, we derived a mean value of $[\alpha/{\rm Fe}]_{\rm spec} = +0.24 \pm 0.07$ with spectroscopic parameters and $[\alpha/{\rm Fe}]_{\rm phot} = +0.23 \pm 0.08$. This $\alpha$-enhancement is consistent with both bulge field stars and with other BGCs, indicating that its chemical enrichment was very similar.

\item We considered Ni as the only iron-peak element with a reliable abundance measurement and find that NGC 6304 is in very good agreement with other BGCs and is also consistent with the bulge field stars.

\item We observed significant star-to-star variation in  C and N ($\sigma_{\rm spec}=$ 0.54, $\sigma_{\rm phot}=$ 0.46). Moreover, NGC 6304 exhibits a clear C-N anticorrelation, observed in both {\ttfamily BACCHUS} and {\ttfamily ASPCAP} results, characterized by a strong enrichment in N and a depletion in C, clearly indicating the presence of MP.

\item We found a significant spread in Na but the variation in O is minimal. This behavior is observed in both {\ttfamily BACCHUS} and {\ttfamily ASPCAP} results. The pattern is very similar to that found in other metal-rich BGCs (\citealp{2017A&A...605A..12M, 2018A&A...620A..96M, 2020MNRAS.492.3742M}; \citealp{2021MNRAS.503.4336M}), suggesting that the formation environment and chemical evolution of metal-rich BGCs may differ from those of typical halo GCs. Additionally, our findings are consistent with previous results from \citet{2009A&A...505..117C}, which highlight the role of cluster mass in shaping the Na-O anticorrelation.

\item No evidence of a Mg-Al anticorrelation was observed for NGC 6304. This absence is consistent with previous observations in other metal-rich BGCs, supporting the idea that the MgAl cycle is less efficient at higher metallicities (\citealp{2017A&A...601A.112P}; \citealp{2020MNRAS.492.1641M}). 

\item A correlation between Ce-N and Ce-Al is observed for NGC 6304, which is present in both {\ttfamily BACCHUS} photometric and {\ttfamily ASPCAP} results. These trends suggest that the enrichment of Ce could be produced by AGB stars (\citealp{2016ApJ...831L..17V}). Similar correlations have been observed before in other metal-rich BGCs (\citealp{2021ApJ...918L...9F, 2022A&A...658A.116F}; \citealp{2025A&A...699A.128B}; \citealp{2025A&A...701A.159F}), positioning Ce as an additional tracer of MP and opening a door to new hypotheses about the origin of MP in GCs. This is indeed very welcome as no single MP scenario to date has been successful at explaining the wide array of observations associated with this fascinating phenomena.

\end{enumerate}

\begin{acknowledgements}
C. Montecinos gratefully acknowledges the support provided by the National Agency for Research and Development (ANID) through the Programa Nacional de Becas de Doctorado (DOCTORADO BECAS CHILE/2022 - 21220138), also gratefully acknowledges the guidance of J. Fernández-Trincado in learning to use the BACCHUS code, and and valuable diskussions with Katia Cunha.
S.V. and D.G. gratefully acknowledge the support provided by Fondecyt Regular n. 1220264 and by the ANID BASAL project FB210003. D.G. also acknowledges financial support from the Direcci\'on de Investigaci\'on y Desarrollo de
la Universidad de La Serena through the Programa de Incentivo a la Investigaci\'on de Acad\'emicos (PIA-DIDULS). C.M. thanks the support provided by ANID-GEMINI Postdoctorado No.32230017.
We thank the referee for their valuable comments and suggestions that helped improve this paper.
\end{acknowledgements}


\begin{appendix}
\onecolumn

\section{Basic parameters of members stars}

\begin{table}[htbp]
 \caption{APOGEE-ID, coordinates, PM, S/N, RVs, 2MASS magnitudes (J, H, and $K_{\mathrm{S}}$) and \textit{Gaia} DR3 magnitudes (G, G$_{\mathrm{BP}}$, and G$_{\mathrm{RP}}$).}
 \label{A1}
 \resizebox{\textwidth}{!}{%
 \begin{tabular}{lllllllllllll}\\
  \hline
  \hline
  \small APOGEE-ID & \small RA$_{\text{J2000}}$ & \small DEC$_{\text{J2000}}$ & \small pmRA & \small pmDEC & \small S/N & \small RV & \small J & \small H & \small $K_{\mathrm{S}}$& \small G & \small G$_{\mathrm{BP}}$ & \small G$_{\mathrm{RP}}$\\
    & \small [deg] & \small [deg] & \small [mas yr$^{-1}$] & \small [mas yr$^{-1}$] & \small [pix$^{-1}$] & \small [km s$^{-1}$] & \small [mag] & \small [mag] & \small [mag] & \small [mag] & \small [mag] & \small [mag]\\
  \hline
  \hline
  \small2M17140389-2929183 & \small258.5162 & \small-29.4884 & \small-4.16 & \small-1.15 & \small55 & \small-110.76 & \small13.57 & \small12.90 & \small12.77 & \small15.81 & \small16.58 & \small14.86 \\[2pt]

  \small2M17141615-2925204 & \small258.5673 & \small-29.4223 & \small-4.03 & \small-1.05 & \small64 & \small-104.02 & \small12.80 & \small11.96 & \small11.82 & \small15.39 & \small16.39 & \small14.39 \\[2pt] 

  \small2M17141646-2926455 & \small258.5686 & \small-29.4459 & \small-4.28 & \small-0.96 & \small51 & \small-106.79 & \small13.43 & \small12.79 & \small12.53 & \small15.77 & \small16.51 & \small14.78 \\[2pt]

  \small2M17142210-2930465 & \small258.5920 & \small-29.5129 & \small3.93 & \small-1.01 & \small66 & \small-111.82 & \small12.24 & \small11.73 & \small11.19 & \small15.12 & \small16.10 & \small14.12 \\[2pt] 
  
  \small2M17142243-2929316 & \small258.5934 & \small-29.4921 & \small-3.77 & \small-1.06 & \small115 & \small-110.40 & \small12.02 & \small11.10 & \small10.94 & \small14.59 & \small15.61 & \small13.58 \\[2pt]

  \small2M17142549-2926198 & \small258.6062 & \small-29.4388 & \small-4.16 & \small-1.21 & \small63 & \small-110.12 & \small13.02 & \small12.19 & \small12.03 & \small15.42 & \small16.40 & \small14.45 \\[2pt]

  \small2M17142716-2929529 & \small258.6131 & \small-29.4980 & \small-4.17 & \small-1.17 & \small78 & \small-106.28 & \small12.54 & \small11.72 & \small11.56 & \small15.11 & \small16.07 & \small14.15 \\[2pt] 
  
  \small2M17142865-2927003 & \small258.6194 & \small-29.4500 & \small-3.93 & \small-1.03 & \small138 & \small-107.23 & \small11.65 & \small10.73 & \small10.52 & \small14.26 & \small15.34 & \small13.23 \\[2pt]

  \small2M17142941-2922414 & \small258.6225 & \small-29.3781 & \small-4.01 & \small-1.04 & \small78 & \small-112.98 & \small13.13 & \small12.36 & \small12.17 & \small15.63 & \small16.48 & \small14.60 \\[2pt]

  \small2M17143081-2927457 & \small258.6284 & \small-29.4627 & \small-4.20 & \small-1.04 & \small117 & \small-109.20 & \small10.50 & \small8.59 & \small8.31 & \small13.61 & \small14.73 & \small12.40 \\[2pt] 

  \small2M17143117-2930149 & \small258.6298 & \small-29.5041 & \small-3.94 & \small-1.31 & \small88 & \small-104.37 & \small12.33 & \small11.46 & \small11.21 & \small14.85 & \small15.88 & \small13.84\\[2pt] 

  \small2M17143158-2932145 & \small258.6316 & \small-29.5373 & \small-4.00 & \small-1.04 & \small60 & \small-108.29 & \small13.63 & \small12.95 & \small12.76 & \small15.88 & \small16.73 & \small14.95\\[2pt] 

  \small2M17143280-2927477 & \small258.6366 & \small-29.4632 & \small-3.88 & \small-1.17 & \small256 & \small-120.59 & \small10.25 & \small9.34 & \small9.00 & \small13.44 & \small14.42 & \small12.15 \\[2pt]

  \small2M17143323-2925484 & \small258.6384 & \small-29.4301 & \small-4.49 & \small-0.91 & \small583 & \textbf{}-115.04 & \small9.01 & \small7.88 & \small7.49 & \small12.92 & \small15.63 & \small11.48 \\[2pt]

  \small2M17143528-2929251 & \small258.6470 & \small-29.4903 & \small-3.94 & -\small1.14 & \small62 & \small-112.26 & \small12.96 & \small12.11 & \small11.93 & \small15.35 & \small16.28 & \small14.37 \\[2pt]

  \small2M17143573-2927050 & \small258.6489 & \small-29.4514 & \small-4.06 & \small-1.12 & \small430 & \small-105.69 & \small9.36 & \small8.36 & \small7.97 & \small12.86 & \small14.49 & \small11.62 \\[2pt]
  
  \small2M17143640-2928513 & \small258.6517 & \small-29.4809 & -\small4.16 & \small-1.24 & \small134 & \small-99.15 & \small11.57 & \small10.68 & \small10.42 & \small14.26 & \small15.33 & \small13.21 \\[2pt] 

  \small2M17143930-2926411 & \small258.6637 & \small-29.4447 & \small-4.12 & \small-1.04 & \small103 & \small-104.31 & \small12.11 & \small11.23 & \small11.04 & \small14.62 & \small15.59 & \small13.60 \\[2pt] 

  \small2M17143986-2924168 & \small258.6660 & \small-29.4046 & \small-4.15 & \small-0.95 & \small58 & \small-107.18 & \small13.65 & \small12.95 & \small12.83 & \small15.83 & \small16.66 & \small14.92 \\[2pt] 
   
  \small2M17144355-2929202 & \small258.6814 & \small-29.4889 & \small-4.08 & \small-1.15 & \small86 & \small-107.90 & \small12.20 & \small11.26 & \small11.06 & \small14.84 & \small15.92 & \small13.81 \\[2pt]

  \small2M17144517-2923296 & \small258.6882 & \small-29.3915 & \small-3.89 & \small-0.99 & \small64 & \small-108.13 & \small13.42 & \small12.75 & \small12.52 & \small15.80 & \small16.59 & \small14.84 \\[2pt] 

  \small2M17144561-2925135 & \small258.6900 & \small-29.4204 & \small-4.12 & \small-1.02 & \small64 & \small-103.74 & \small13.53 & \small12.73 & \small12.57 & \small15.77 & \small16.67 & \small14.82 \\[2pt] 

  \small2M17144732-2927090 & \small258.6971 & \small-29.4525 & \small-3.89 & \small-1.26 & 57 & \small-109.72 & \small13.62 & \small12.90 & \small12.72 & \small16.04 & \small16.95 & \small15.07 \\[2pt] 

  \small2M17144851-2930108 & \small258.7021 & \small-29.5030 & \small-4.03 & \small-1.11 & \small51 & \small-109.25 & \small13.42 & \small12.83 & \small12.65 & \small15.88 & \small16.71 & \small14.92 \\[2pt] 

  \small2M17144988-2928301 & \small258.7078 & \small-29.4750 & \small-4.03 & \small-1.14 & \small64 & \small-107.43 & \small12.96 & \small12.16 & \small11.97 & \small15.69 & \small16.45 & \small14.46 \\[2pt] 

  \small2M17145331-2925511 & \small258.7221 & \small-29.4308 & \small-4.09 & \small-0.94 & \small60 & \small-107.21 & \small13.07 & \small12.34 & \small12.18 & \small15.98 & \small16.50 & \small14.67 \\[2pt] 

  \small2M17145815-2927517 & \small258.7423 & \small-29.4643 & \small-4.19 & \small-1.10 & \small68 & \small-109.54 & \small12.92 & \small12.11 & \small11.91 & \small15.44 & \small16.43 & \small14.45 \\[2pt] 

  \hline
 \end{tabular}
 }
\end{table}

\clearpage
\section{Spectroscopic and photometric stellar parameters}

\begin{table}[htbp]
\centering
 \caption{Spectroscopic and photometric stellar parameters.}
 \label{B2}
\resizebox{18cm}{!} {
\begin{tabular}{llllllllllllr}
\hline
\hline
 \multicolumn{5}{r}{\small Spectroscopic parameters} & \multicolumn{7}{r}{\small Photometric parameters}\\
\cline{3-6}
\cline{10-13}
\small APOGEE-ID & & \small $T_{\mathrm{eff}}$ & \small $\log(g)$ & \small $\xi_{\mathrm{t}}$ & \small [Fe/H] & & & & \small $T_{\mathrm{eff}}$ & \small $\log(g)$ & \small $\xi_{\mathrm{t}}$ & [\small Fe/H]\\
& & \small [K] & \small [dex] & \small [km s$^{-1}$] & \small [dex]  & & & & \small [K] & \small [dex] & \small [km s$^{-1}$] & \small [dex]\\
\hline
\hline
  \small2M17140389-2929183 & & \small4935 & \small2.35 & \small2.04 & \small-0.51 & & & & \small4598 & \small2.35 & \small1.15 & \small-0.55\\
  
  \small2M17141615-2925204 & & \small4935 & \small1.95 & \small1.44 & \small-0.51 & & & & \small4382 & \small1.93 & \small1.21 & \small-0.28\\
  
  \small2M17141646-2926455 & & \small4712 & \small2.50 & \small1.31 & \small-0.51 & & & & \small4550 & \small2.26 & \small1.17 & \small-0.52\\ 

  \small2M17142210-2930465 & & \small4335 & \small2.05 & \small1.43 & \small-0.47 & & & & \small4238 & \small1.66 & \small1.25 & \small-0.48\\

  \small2M17142243-2929316 & & \small4262 & \small2.01 & \small1.73 & \small-0.46 & & & & \small4180 & \small1.55 & \small1.27 & \small-0.41\\
  
  \small2M17142549-2926198 & & \small4276 & \small2.19 & \small1.35 & \small-0.32 & & & & \small4434 & \small2.03 & \small1.20 & \small-0.17\\  

 \small 2M17142716-2929529 & & \small4388 & \small2.06 & \small1.67 & \small-0.43 & & & & \small4327 & \small1.82 & \small1.23 & \small-0.37\\ 

  \small2M17142865-2927003 & & \small4225 & \small1.79 & \small1.83 & \small-0.47 & & & & \small4080 & \small1.36 & \small1.29 & \small-0.54\\ 
  
  \small2M17142941-2922414 & & \small4536 & \small2.36 & \small1.51 & \small-0.52 & & & & \small4465 & \small2.09 & \small1.19 & \small-0.47\\ 

  \small2M17143081-2927457 & & \small3787 & \small1.22 & \small1.76 & \small-0.53 & & & & \small3526 & \small0.38 & \small1.41 & \small-0.48\\ 

  \small2M17143117-2930149 & & \small4118 & \small1.81 & \small1.36 & \small-0.41 & & & & \small4241 & \small1.66 & \small1.25 & \small-0.36\\ 

  \small2M17143158-2932145 & & \small4834 & \small2.40 & \small1.70 & \small-0.55 & & & & \small4595 & \small2.35 & \small1.15 & \small-0.56\\ 

  \small2M17143280-2927477 & & \small3790 & \small1.24 & \small1.93 & \small-0.61 & & & & \small3707 & \small0.69 & \small1.38 & \small-0.32\\  

  \small2M17143323-2925484 & & \small3503 & \small1.25 & \small2.31 & \small-0.62 & & & & \small3297 & \small-0.01 & \small1.45 & \small-0.47\\  

  \small2M17143528-2929251 & & \small4471 & \small2.27 & \small1.57 & \small-0.49 & & & & \small4412 & \small1.98 & \small1.21 & \small-0.46\\ 

  \small2M17143573-2927050 & & \small3812 & \small1.34 & \small2.66 & \small-0.47 & & & & \small3425 & \small0.21 & \small1.42 & \small-0.45\\  

  \small2M17143640-2928513 & & \small3959 & \small1.50 & \small1.68 & \small-0.44 & & & & \small4061 & \small1.33 & \small1.30 & \small-0.32\\ 
  
  \small2M17143930-2926411 & & \small4150 & \small1.67 & \small1.52 & \small-0.47 & & & & \small4199 & \small1.58 & \small1.26 & \small-0.39\\  

  \small2M17143986-2924168 & & \small4950 & \small2.49 & \small1.61 & \small-0.38 & & & & \small4612 & \small2.38 & \small1.15 & \small-0.49\\  

  \small2M17144355-2929202 & & \small4324 & \small2.11 & \small1.52 & \small-0.45 & & & & \small4201 & \small1.59 & \small1.26 & \small-0.36\\ 

  \small2M17144517-2923296 & & \small4795 & \small2.39 & \small1.58 & \small-0.58 & & & & \small4545 & \small2.24 & \small1.17 & \small-0.45\\   

  \small2M17144561-2925135 & & \small4654 & \small2.29 & \small1.60 & \small-0.37 & & & & \small4552 & \small2.26 & \small1.17 & \small-0.38\\   

  \small2M17144732-2927090 & & \small4775 & \small2.48 & \small1.48 & \small-0.50 & & & & \small4582 & \small2.32 & \small1.16 & \small-0.55\\  

  \small2M17144851-2930108 & & \small5010 & \small2.55 & \small1.94 & \small-0.42 & & & & \small4569 & \small2.29 & \small1.16 & \small-0.44\\   

  \small2M17144988-2928301 & & \small4365 & \small2.09 & \small1.07 & \small-0.52 & & & & \small4413 & \small1.99 & \small1.21 & \small-0.55\\   

  \small2M17145331-2925511 & & \small5035 & \small2.64 & \small1.52 & \small-0.49 & & & & \small4459 & \small2.08 & \small1.19 & \small-0.74\\  
  
  \small2M17145815-2927517 & & \small4468 & \small2.27 & \small1.58 & \small-0.47 & & & & \small4405 & \small1.97 & \small1.21 & \small-0.41\\
 \hline
\end{tabular}
}
\end{table}

\clearpage
\section{Abundance for NGC 6304}

\begin{table}[htbp]
\centering
\caption{{\ttfamily BACCHUS} elemental abundances of the observed stars using spectroscopic parameters.}
\label{tab:C1}
\resizebox{\textwidth}{5cm}{%
\renewcommand{\arraystretch}{3.5}
\begin{tabular}{lcc|cccccccc|cccccccccc}
 \hline       
 \hline
 \multicolumn{3}{l}{\Huge \textbf{Spectroscopy}} &  \multicolumn{8}{|c|}{\Huge \textbf{Good}} & 
 \multicolumn{10}{c}{\Huge \textbf{Intermediate}}\\[2pt]
 \hline
 \hline 
 \\  
 \Huge APOGEE-ID & \Huge[$\alpha$/Fe] &  & \Huge[C/Fe] & \Huge[O/Fe] & \Huge[Mg/Fe & \Huge[Al/Fe] & \Huge[Si/Fe] & \Huge[Ca/Fe] & \Huge[Fe/H] & \Huge[Ni/Fe] &  & \Huge[N/Fe] & \Huge[Na/Fe] & \Huge[S/Fe] & \Huge[K/Fe] & \Huge[Ti/Fe] & \Huge[V/Fe] & \Huge[Cr/Fe] & \Huge[Mn/Fe] & \Huge[Ce/Fe]\\
 \hline 
 \hline     
 \\  
\Huge2M17140389-2929183 & \Huge0.29 &  & \Huge-0.82 & \Huge {...} & \Huge0.30 & \Huge0.40 & \Huge0.30 & \Huge0.26 & \Huge-0.53 & \Huge0.02 &  & \Huge1.56 & \Huge {...} & \Huge0.37 & \Huge {...} & \Huge0.52 & \Huge {...} & \Huge0.09 & \Huge0.17 & \Huge {...} \\[2pt]
  
\Huge2M17141615-2925204 & \Huge0.15 &  & \Huge-0.10 & \Huge0.10 & \Huge0.27 & \Huge0.20 & \Huge0.13 & \Huge0.05 & \Huge-0.38 & \Huge-0.06 &  & \Huge-0.05 & \Huge0.01 & \Huge0.52 & \Huge0.21 & \Huge-0.07 & \Huge-0.21 & \Huge0.09 & \Huge-0.27 & \Huge-0.10 \\[2pt]

\Huge2M17141646-2926455 & \Huge0.33 &  & \Huge0.10 & \Huge0.34 & \Huge0.31 & \Huge0.28 & \Huge0.38 & \Huge0.31 & \Huge-0.52 & \Huge0.22 &  & \Huge0.68 & \Huge {...} & \Huge1.28 & \Huge0.66 & \Huge1.09 & \Huge {...} & \Huge0.02  & \Huge {...} & \Huge {...} \\[2pt]

\Huge2M17142210-2930465 & \Huge0.34 &  & \Huge-0.24 & \Huge0.17 & \Huge0.39 & \Huge0.31 & \Huge0.44 & \Huge0.20 & \Huge-0.46 & \Huge0.26 &  & \Huge0.96 & \Huge0.34 & \Huge {...} & \Huge0.32 & \Huge0.28 & \Huge {...} & \Huge0.15  & \Huge0.02 & \Huge0.08 \\[2pt]
  
\Huge2M17142243-2929316 & \Huge0.28 &  & \Huge-0.38 & \Huge0.11 & \Huge0.38 & \Huge0.23 & \Huge0.31 & \Huge0.14 & \Huge-0.41 & \Huge0.13 &  & \Huge1.08 & \Huge0.49 & \Huge1.22 & \Huge0.61 & \Huge0.40 & \Huge0.19 & \Huge0.17  & \Huge {...} & \Huge0.07 \\[2pt]

\Huge2M17142549-2926198 & \Huge0.31 &  & \Huge0.04 & \Huge0.17 & \Huge0.44 & \Huge0.26 & \Huge0.37 & \Huge0.11 & \Huge-0.38 & \Huge0.14 &  & \Huge-0.05 & \Huge0.17 & \Huge0.48 & \Huge0.31 & \Huge-0.03 & \Huge0.00 & \Huge0.13  & \Huge-0.08 & \Huge0.05 \\[2pt]

\Huge2M17142716-2929529 & \Huge0.27 &  & \Huge-0.34 & \Huge0.13 & \Huge0.31 & \Huge0.42 & \Huge0.27 & \Huge0.23 & \Huge-0.48 & \Huge0.15 &  & \Huge {...} & \Huge0.27 & \Huge0.70 & \Huge0.61 & \Huge0.73 & \Huge0.26 & \Huge0.18  & \Huge0.00 & \Huge0.00 \\[2pt]
  
\Huge2M17142865-2927003 & \Huge0.31 &  & \Huge-0.56 & \Huge0.14 & \Huge0.43 & \Huge0.49 & \Huge0.30 & \Huge0.19 & \Huge-0.47 & \Huge0.16 &  & \Huge1.24 & \Huge {...} & \Huge0.36 & \Huge0.63 & \Huge0.40 & \Huge0.28 & \Huge0.30  & \Huge0.12 & \Huge0.09 \\[2pt]

\Huge2M17142941-2922414 & \Huge0.26 &  & \Huge-0.20 & \Huge0.15 & \Huge0.27 & \Huge0.39 & \Huge0.34 & \Huge0.16  & \Huge-0.46 & \Huge0.17 &  & \Huge0.87 & \Huge0.41 & \Huge1.08 & \Huge0.62 & \Huge {...} & \Huge0.13 & \Huge0.24  & \Huge0.08 & \Huge-0.17 \\[2pt]

\Huge2M17143081-2927457 & \Huge0.30 &  & \Huge0.30 & \Huge0.31 & \Huge0.38 & \Huge0.26 & \Huge0.33 & \Huge0.18 & \Huge-0.46 & \Huge0.14 &  & \Huge-0.11 & \Huge {...} & \Huge0.44 & \Huge0.30 & \Huge0.25 & \Huge0.20 & \Huge {...}  & \Huge-0.04 & \Huge0.00 \\[2pt]

\Huge2M17143117-2930149 & \Huge0.21 &  & \Huge0.08 & \Huge0.20 & \Huge0.21 & \Huge0.30 & \Huge0.31 & \Huge0.10 & \Huge-0.39 & \Huge0.10 &  & \Huge-0.01 & \Huge0.09 & \Huge0.49 & \Huge0.20 & \Huge0.18 & \Huge0.21 & \Huge-0.07  & \Huge-0.13 & \Huge-0.20 \\[2pt]

\Huge2M17143158-2932145 & \Huge0.19 &  & \Huge-0.06 & \Huge0.43 & \Huge0.24 & \Huge0.20 & \Huge0.24 & \Huge0.10 & \Huge-0.46 & \Huge0.03 &  & \Huge0.82 & \Huge {...} & \Huge0.18 & \Huge0.36 & \Huge0.22 & \Huge {...} & \Huge0.65  & \Huge {...} & \Huge {...} \\[2pt]
  
\Huge2M17143280-2927477 & \Huge0.21 &  & \Huge0.01 & \Huge0.16 & \Huge0.31 & \Huge0.16 & \Huge0.28 & \Huge0.03 & \Huge-0.42 & \Huge0.05 &  & \Huge0.51 & \Huge {...} & \Huge {...} & \Huge0.50 & \Huge0.10 & \Huge0.13 & \Huge0.12  & \Huge {...} & \Huge-0.02 \\[2pt]

\Huge2M17143323-2925484 & \Huge0.18 &  & \Huge-0.11 & \Huge0.18 & \Huge0.21 & \Huge {...} & \Huge0.26 & \Huge0.07 & \Huge-0.52 & \Huge0.16 &  & \Huge0.68 & \Huge0.60 & \Huge {...} & \Huge0.19 & \Huge0.26 & \Huge0.26 & \Huge {...}  & \Huge0.08 & \Huge0.15 \\[2pt]

\Huge2M17143528-2929251 & \Huge0.22 &  & \Huge-0.53 & \Huge0.11 & \Huge0.24 & \Huge0.33 & \Huge0.23 & \Huge0.19 & \Huge-0.45 & \Huge0.12 &  & \Huge1.21 & \Huge {...} & \Huge0.22 & \Huge0.52 & \Huge0.78 & \Huge0.12 & \Huge0.25  & \Huge0.55 & \Huge {...} \\[2pt]

\Huge2M17143573-2927050 & \Huge0.06 &  & \Huge-0.50 & \Huge0.29 & \Huge0.14 & \Huge {...} & \Huge0.10 & \Huge-0.08 & \Huge-0.48 & \Huge0.05 &  & \Huge1.32 & \Huge0.34 & \Huge0.31 & \Huge0.09 & \Huge-0.01 & \Huge0.16 & \Huge-0.11  & \Huge-0.02 & \Huge0.01 \\[2pt]
  
\Huge2M17143640-2928513 & \Huge0.26 &  & \Huge0.10 & \Huge0.21 & \Huge0.41 & \Huge0.22 & \Huge0.29 & \Huge0.08 & \Huge-0.41 & \Huge0.08 &  & \Huge0.02 & \Huge {...} & \Huge0.47 & \Huge0.26 & \Huge0.22 & \Huge0.02 & \Huge-0.03  & \Huge-0.05 & \Huge-0.13 \\[2pt] 

\Huge2M17143930-2926411 & \Huge0.17 &  & \Huge0.02 & \Huge0.12 & \Huge0.29 & \Huge0.09 & \Huge0.23 & \Huge-0.01 & \Huge-0.37 & \Huge-0.03 &  & \Huge-0.08 & \Huge0.05 & \Huge0.19 & \Huge0.16 & \Huge-0.02 & \Huge-0.06 & \Huge-0.08  & \Huge-0.18 & \Huge-0.13 \\[2pt] 

\Huge2M17143986-2924168 & \Huge0.15 &  & \Huge {...} & \Huge {...} & \Huge0.14 & \Huge0.24 & \Huge0.24 & \Huge0.07 & \Huge-0.40 & \Huge-0.01 &  & \Huge1.51 & \Huge0.06 & \Huge0.28 & \Huge0.34 & \Huge0.96 & \Huge {...} & \Huge-0.16  & \Huge-0.07 & \Huge-0.02 \\[2pt] 
   
\Huge2M17144355-2929202 & \Huge0.22 &  & \Huge-0.35 & \Huge0.17 & \Huge0.20 & \Huge0.37 & \Huge0.28 & \Huge0.16 & \Huge-0.46 & \Huge0.16 &  & \Huge1.09 & \Huge0.71 & \Huge0.93 & \Huge0.66 & \Huge0.52 & \Huge0.24 & \Huge0.29  & \Huge {...} & \Huge0.19 \\[2pt]

\Huge2M17144517-2923296 & \Huge0.30 &  & \Huge0.16 & \Huge0.28 & \Huge0.32 & \Huge0.31 & \Huge0.34 & \Huge0.25 & \Huge-0.50 & \Huge0.08 &  & \Huge0.13 & \Huge0.18 & \Huge0.38 & \Huge {...} & \Huge0.33 & \Huge {...} & \Huge0.76  & \Huge-0.11 & \Huge0.10 \\[2pt]

\Huge2M17144561-2925135 & \Huge0.36 &  & \Huge-0.04 & \Huge0.43 & \Huge0.38 & \Huge0.32 & \Huge0.42 & \Huge0.28 & \Huge-0.48 & \Huge0.16 &  & \Huge0.75 & \Huge {...} & \Huge0.50 & \Huge0.37 & \Huge0.33 & \Huge {...} & \Huge-0.21  & \Huge-0.06 & \Huge {...} \\[2pt] 

\Huge2M17144732-2927090 & \Huge0.25 &  & \Huge0.38 & \Huge0.40 & \Huge0.40 & \Huge0.26 & \Huge0.30 & \Huge0.06 & \Huge-0.52 & \Huge0.16 &  & \Huge-0.40 & \Huge {...} & \Huge0.37 & \Huge0.26 & \Huge0.27 & \Huge {...} & \Huge0.36  & \Huge-0.16 & \Huge-0.01 \\[2pt] 

\Huge2M17144851-2930108 & \Huge0.20 &  & \Huge {...} & \Huge {...} & \Huge0.13 & \Huge0.22 & \Huge0.30 & \Huge0.18 & \Huge-0.43 & \Huge-0.07 &  & \Huge0.98 & \Huge {...} & \Huge0.34 & \Huge0.27 & \Huge0.50 & \Huge {...} & \Huge {...} & \Huge {...} & \Huge {...} \\[2pt]

\Huge2M17144988-2928301 & \Huge0.23 &  & \Huge-0.09 & \Huge0.23 & \Huge0.26 & \Huge0.23 & \Huge0.27 & \Huge0.17 & \Huge-0.51 & \Huge0.06 &  & \Huge0.81 & \Huge0.24 & \Huge0.98 & \Huge0.56 & \Huge0.36 & \Huge0.25 & \Huge {...}  & \Huge-0.00 & \Huge0.10 \\[2pt]

\Huge2M17145331-2925511 & \Huge0.22 &  & \Huge {...} & \Huge0.48 & \Huge0.18 & \Huge0.29 & \Huge0.26 & \Huge0.23 & \Huge-0.42 & \Huge0.08 &  & \Huge0.63 & \Huge0.23 & \Huge0.21 & \Huge0.25 & \Huge0.50 & \Huge {...} & \Huge0.20  & \Huge {...} & \Huge {...} \\[2pt] 

\Huge2M17145815-2927517 & \Huge0.24 &  & \Huge-0.16 & \Huge0.20 & \Huge0.31 & \Huge0.25 & \Huge0.22 & \Huge0.19 & \Huge-0.43 & \Huge0.07 &  & \Huge0.74 & \Huge {...} & \Huge0.53 & \Huge0.59 & \Huge0.54 & \Huge {...} & \Huge0.07  & \Huge-0.01 & \Huge0.10 \\[2pt]
\\
 \hline 
\Huge \textbf{Cluster} & \Huge0.24$\pm$0.01 &  & \Huge-0.14$\pm$0.06 & \Huge0.23$\pm$0.02 & \Huge0.29$\pm$0.02 & \Huge0.28$\pm$0.02 & \Huge0.29$\pm$0.01 & \Huge0.14$\pm$0.02 & \Huge-0.45$\pm$0.01 & \Huge0.10$\pm$0.02 &  & \Huge0.65$\pm$0.11 & \Huge0.28$\pm$0.05 & \Huge0.54$\pm$0.07 & \Huge0.39$\pm$0.04 & \Huge0.37$\pm$0.06 & \Huge0.14$\pm$0.03 & \Huge0.15$\pm0.05$ & \Huge-0.01$\pm0.04$ & \Huge0.01$\pm$0.02\\[8pt]
 \hline
 \hline 
 \end{tabular}
}
\end{table}

\begin{table}[htbp]
\centering
\caption{{\ttfamily BACCHUS} elemental abundances of the observed stars using photometric parameters.}
\label{tab:C2}
\resizebox{\textwidth}{5cm}{%
\renewcommand{\arraystretch}{3.5}
\begin{tabular}{lcc|cccccccc|cccccccccc}
 \hline       
 \hline
 \multicolumn{3}{l}{\Huge \textbf{Photometry}} &  \multicolumn{8}{|c|}{\Huge \textbf{Good}} & 
 \multicolumn{10}{c}{\Huge \textbf{Intermediate}}\\[2pt]
 \hline
 \hline 
 \\  
 \Huge APOGEE-ID & \Huge[$\alpha$/Fe] &  & \Huge[C/Fe] & \Huge[O/Fe] & \Huge[Mg/Fe & \Huge[Al/Fe] & \Huge[Si/Fe] & \Huge[Ca/Fe] & \Huge[Fe/H] & \Huge[Ni/Fe] &  & \Huge[N/Fe] & \Huge[Na/Fe] & \Huge[S/Fe] & \Huge[K/Fe] & \Huge[Ti/Fe] & \Huge[V/Fe] & \Huge[Cr/Fe] & \Huge[Mn/Fe] & \Huge[Ce/Fe]\\
 \hline 
 \hline     
 \\  
\Huge2M17140389-2929183 & \Huge0.20 &  & \Huge {...} & \Huge {...} & \Huge0.16 & \Huge0.22 & \Huge0.36 & \Huge0.06 & \Huge-0.52 & \Huge0.08 &  & \Huge1.04 & \Huge0.26 & \Huge0.42 & \Huge {...} & \Huge0.24 & \Huge {...} & \Huge-0.26 & \Huge-0.04 & \Huge {...} \\[2pt]
  
\Huge2M17141615-2925204 & \Huge0.24 &  & \Huge-0.02 & \Huge0.26 & \Huge0.35 & \Huge0.37 & \Huge0.23 & \Huge0.16 & \Huge-0.35 & \Huge0.05 &  & \Huge0.32 & \Huge0.09 & \Huge0.44 & \Huge0.37 & \Huge0.19 & \Huge0.12 & \Huge0.30 & \Huge-0.19 & \Huge0.08 \\[2pt]

\Huge2M17141646-2926455 & \Huge0.23 &  & \Huge-0.12 & \Huge0.32 & \Huge0.25 & \Huge0.19 & \Huge0.24 & \Huge0.21 & \Huge-0.47 & \Huge0.16 &  & \Huge0.57 & \Huge {...} & \Huge1.23 & \Huge0.56 & \Huge0.80 & \Huge {...} & \Huge-0.15 & \Huge {...} & \Huge {...} \\[2pt]

\Huge2M17142210-2930465 & \Huge0.28 &  & \Huge-0.32 & \Huge0.10 & \Huge0.38 & \Huge0.31 & \Huge0.32 & \Huge0.14 & \Huge-0.48 & \Huge0.18 &  & \Huge0.86 & \Huge0.25 & \Huge0.74 & \Huge0.27 & \Huge0.20 & \Huge {...} & \Huge0.04 & \Huge-0.08 & \Huge-0.10 \\[2pt]
  
\Huge2M17142243-2929316 & \Huge0.28 &  & \Huge-0.49 & \Huge-0.01 & \Huge0.34 & \Huge0.37 & \Huge0.36 & \Huge0.14 & \Huge-0.42 & \Huge0.07 &  & \Huge1.06 & \Huge {...} & \Huge0.72 & \Huge0.55 & \Huge0.56 & \Huge0.12 & \Huge0.13 & \Huge-0.03 & \Huge-0.14 \\[2pt]

\Huge2M17142549-2926198 & \Huge0.33 &  & \Huge-0.02 & \Huge0.30 & \Huge0.45 & \Huge0.45 & \Huge0.32 & \Huge0.21 & \Huge-0.31 & \Huge0.14 &  & \Huge {...} & \Huge0.18 & \Huge0.32 & \Huge0.25 & \Huge {...} & \Huge0.26 & \Huge0.09 & \Huge-0.15 & \Huge-0.03 \\[2pt]

\Huge2M17142716-2929529 & \Huge0.29 &  & \Huge-0.45 & \Huge0.09 & \Huge0.38 & \Huge0.50 & \Huge0.32 & \Huge0.18 & \Huge-0.46 & \Huge0.11 &  & \Huge {...} & \Huge0.22 & \Huge0.31 & \Huge0.59 & \Huge0.51 & \Huge0.20 & \Huge0.10 & \Huge-0.09 & \Huge-0.09 \\[2pt]
  
\Huge2M17142865-2927003 & \Huge0.22 &  & \Huge {...} & \Huge0.01 & \Huge0.34 & \Huge0.48 & \Huge0.27 & \Huge0.06 & \Huge-0.48 & \Huge0.03 &  & \Huge0.03 & \Huge {...} & \Huge0.47 & \Huge0.50 & \Huge0.33 & \Huge0.08 & \Huge0.11 & \Huge-0.02 & \Huge {...} \\[2pt]

\Huge2M17142941-2922414 & \Huge0.24 &  & \Huge-0.33 & \Huge0.12 & \Huge0.29 & \Huge0.36 & \Huge0.31 & \Huge0.11 & \Huge-0.44 & \Huge0.13 &  & \Huge0.98 & \Huge0.24 & \Huge0.19 & \Huge0.54 & \Huge0.40 & \Huge0.08 & \Huge0.13 & \Huge-0.03 & \Huge-0.21 \\[2pt]

\Huge2M17143081-2927457 & \Huge0.30 &  & \Huge-0.04 & \Huge-0.01 & \Huge0.48 & \Huge {...} & \Huge {...} & \Huge0.11 & \Huge-0.52 & \Huge-0.06 &  & \Huge-0.03 & \Huge0.26 & \Huge {...} & \Huge0.26 & \Huge-0.00 & \Huge-0.32 & \Huge-0.23 & \Huge-0.10 & \Huge-0.20 \\[2pt]

\Huge2M17143117-2930149 & \Huge0.34 &  & \Huge0.06 & \Huge0.32 & \Huge0.35 & \Huge0.58 & \Huge0.39 & \Huge0.27 & \Huge-0.41 & \Huge0.08 &  & \Huge0.22 & \Huge0.17 & \Huge0.34 & \Huge0.24 & \Huge0.34 & \Huge0.38 & \Huge0.00 & \Huge-0.12 & \Huge-0.16 \\[2pt]

\Huge2M17143158-2932145 & \Huge0.23 &  & \Huge0.04 & \Huge0.29 & \Huge0.23 & \Huge0.15 & \Huge0.41 & \Huge0.04 & \Huge-0.50 & \Huge0.12 &  & \Huge0.50 & \Huge {...} & \Huge0.34 & \Huge0.25 & \Huge0.22 & \Huge {...} & \Huge {...} & \Huge {...} & \Huge {...} \\[2pt]
  
\Huge2M17143280-2927477 & \Huge0.29 &  & \Huge-0.30 & \Huge0.14 & \Huge0.48 & \Huge0.39 & \Huge0.34 & \Huge0.07 & \Huge-0.42 & \Huge0.07 &  & \Huge1.02 & \Huge {...} & \Huge {...} & \Huge0.50 & \Huge0.26 & \Huge0.02 & \Huge0.08 & \Huge0.08 & \Huge {...} \\[2pt]

\Huge2M17143323-2925484 & \Huge0.41 &  & \Huge {...} & \Huge0.13 & \Huge0.43 & \Huge0.15 & \Huge0.62 & \Huge0.17 & \Huge-0.50 & \Huge0.10 &  & \Huge0.13 & \Huge0.52 & \Huge {...} & \Huge0.27 & \Huge0.31 & \Huge0.15 & \Huge-0.15 & \Huge-0.21 & \Huge-0.15 \\[2pt]

\Huge2M17143528-2929251 & \Huge0.27 &  & \Huge-0.58 & \Huge-0.01 & \Huge0.36 & \Huge0.36 & \Huge0.25 & \Huge0.18 & \Huge-0.45 & \Huge0.07 &  & \Huge1.18 & \Huge {...} & \Huge {...} & \Huge0.48 & \Huge {...} & \Huge0.07 & \Huge0.21 & \Huge0.06 & \Huge-0.24 \\[2pt]

\Huge2M17143573-2927050 & \Huge0.07 &  & \Huge {...} & \Huge0.16 & \Huge0.27 & \Huge0.13 & \Huge {...} & \Huge-0.13 & \Huge-0.45 & \Huge0.04 &  & \Huge0.08 & \Huge0.21 & \Huge0.61 & \Huge0.63 & \Huge-0.29 & \Huge-0.22 & \Huge {...} & \Huge0.02 & \Huge {...} \\[2pt]
  
\Huge2M17143640-2928513 & \Huge0.15 &  & \Huge-0.08 & \Huge0.27 & \Huge0.29 & \Huge0.29 & \Huge0.11 & \Huge0.05 & \Huge-0.32 & \Huge0.05 &  & \Huge0.17 & \Huge {...} & \Huge0.05 & \Huge0.06 & \Huge0.18 & \Huge0.00 & \Huge-0.26 & \Huge-0.26 & \Huge-0.32 \\[2pt]

\Huge2M17143930-2926411 & \Huge0.13 &  & \Huge-0.07 & \Huge0.12 & \Huge0.20 & \Huge0.14 & \Huge0.20 & \Huge-0.02 & \Huge-0.29 & \Huge-0.07 &  & \Huge-0.08 & \Huge0.00 & \Huge0.18 & \Huge0.10 & \Huge0.07 & \Huge-0.06 & \Huge {...} & \Huge-0.26 & \Huge-0.42 \\[2pt]

\Huge2M17143986-2924168 & \Huge0.16 &  & \Huge {...} & \Huge {...} & \Huge0.16 & \Huge0.20 & \Huge0.31 & \Huge-0.00 & \Huge-0.50 & \Huge0.01 &  & \Huge1.17 & \Huge {...} & \Huge0.43 & \Huge0.29 & \Huge0.01 & \Huge0.27 & \Huge0.04 & \Huge-0.18 & \Huge {...} \\[2pt]
   
\Huge2M17144355-2929202 & \Huge0.24 &  & \Huge-0.55 & \Huge0.09 & \Huge0.28 & \Huge0.47 & \Huge0.33 & \Huge0.10 & \Huge-0.50 & \Huge0.13 &  & \Huge {...} & \Huge0.59 & \Huge0.09 & \Huge0.58 & \Huge0.50 & \Huge0.19 & \Huge0.20 & \Huge-0.02 & \Huge0.01 \\[2pt]

\Huge2M17144517-2923296 & \Huge0.07 &  & \Huge-0.03 & \Huge0.06 & \Huge0.10 & \Huge0.10 & \Huge0.17 & \Huge-0.05 & \Huge-0.42 & \Huge-0.07 &  & \Huge-0.13 & \Huge {...} & \Huge0.19 & \Huge0.03 & \Huge-0.43 & \Huge0.12 & \Huge-0.12 & \Huge-0.45 & \Huge-0.27 \\[2pt]

\Huge2M17144561-2925135 & \Huge0.15 &  & \Huge-0.10 & \Huge0.19 & \Huge0.15 & \Huge0.06 & \Huge0.24 & \Huge0.06 & \Huge-0.37 & \Huge-0.02 &  & \Huge0.41 & \Huge {...} & \Huge0.27 & \Huge0.08 & \Huge-0.02 & \Huge {...} & \Huge-0.14 & \Huge-0.16 & \Huge {...} \\[2pt]

\Huge2M17144732-2927090 & \Huge0.22 &  & \Huge0.36 & \Huge0.08 & \Huge0.37 & \Huge0.22 & \Huge0.31 & \Huge-0.01 & \Huge-0.57 & \Huge0.14 &  & \Huge-0.36 & \Huge {...} & \Huge0.51 & \Huge0.16 & \Huge0.03 & \Huge {...} & \Huge0.17 & \Huge-0.26 & \Huge-0.30 \\[2pt]

\Huge2M17144851-2930108 & \Huge0.18 &  & \Huge-0.32 & \Huge0.28 & \Huge0.08 & \Huge0.24 & \Huge0.43 & \Huge0.05 & \Huge-0.52 & \Huge0.00 &  & \Huge1.05 & \Huge0.59 & \Huge0.41 & \Huge0.14 & \Huge {...} & \Huge {...} & \Huge0.35 & \Huge {...} & \Huge0.27 \\[2pt]

\Huge2M17144988-2928301 & \Huge0.27 &  & \Huge-0.10 & \Huge0.30 & \Huge0.31 & \Huge0.29 & \Huge0.26 & \Huge0.23 & \Huge-0.53 & \Huge0.06 &  & \Huge0.89 & \Huge0.27 & \Huge0.77 & \Huge0.57 & \Huge0.42 & \Huge0.29 & \Huge {...} & \Huge0.12 & \Huge0.07 \\[2pt]

\Huge2M17145331-2925511 & \Huge0.19 &  & \Huge {...} & \Huge {...} & \Huge0.15 & \Huge0.06 & \Huge0.32 & \Huge0.10 & \Huge-0.59 & \Huge0.01 &  & \Huge0.52 & \Huge {...} & \Huge0.86 & \Huge0.13 & \Huge-0.09 & \Huge {...} & \Huge-0.52 & \Huge-0.30 & \Huge {...} \\[2pt]

\Huge2M17145815-2927517 & \Huge0.24 &  & \Huge-0.22 & \Huge-0.01 & \Huge0.26 & \Huge0.26 & \Huge0.27 & \Huge0.17 & \Huge- & \Huge0.03 &  & \Huge0.68 & \Huge {...} & \Huge0.25 & \Huge0.53 & \Huge0.34 & \Huge {...} & \Huge-0.07 & \Huge0.02 & \Huge-0.09 \\[2pt]
\\
 \hline 
\Huge \textbf{Cluster} & \Huge0.23$\pm$0.02 &  & \Huge-0.18$\pm0.05$ & \Huge0.15$\pm0.02$ & \Huge0.29$\pm0.02$ & \Huge0.28$\pm0.03$ & \Huge0.31$\pm0.02$ & \Huge0.10$\pm0.02$ & \Huge-0.45$\pm0.02$ & \Huge0.06$\pm0.01$ &  & \Huge0.51$\pm0.10$ & \Huge0.28$\pm0.04$ & \Huge0.44$\pm0.06$ & \Huge0.34$\pm0.04$ & \Huge0.21$\pm0.06$ & \Huge0.10$\pm0.04$ & \Huge0.01$\pm0.04$ & \Huge-0.11$\pm0.03$ & \Huge-0.13$\pm0.04$\\[8pt]
 \hline
 \hline 
 \end{tabular}
}
\end{table}

\clearpage

\begin{table}[!htbp]
\centering
 \caption{{\ttfamily ASPCAP} elemental abundances of the observed stars.}
 \label{tab:C3}
\resizebox{\textwidth}{5cm}{%
\renewcommand{\arraystretch}{3.5}
\begin{tabular}{lcc|cccccccc|cccccccccc}
 \hline       
 \hline
 \multicolumn{3}{l}{\Huge \textbf{ASPCAP}} &  \multicolumn{8}{|c|}{\Huge \textbf{Good}} & 
 \multicolumn{10}{c}{\Huge \textbf{Intermediate}}\\[2pt]
 \hline
 \hline 
 \\  
 \Huge APOGEE-ID & \Huge[$\alpha$/Fe] &  & \Huge[C/Fe] & \Huge[O/Fe] & \Huge[Mg/Fe & \Huge[Al/Fe] & \Huge[Si/Fe] & \Huge[Ca/Fe] & \Huge[Fe/H] & \Huge[Ni/Fe] &  & \Huge[N/Fe] & \Huge[Na/Fe] & \Huge[S/Fe] & \Huge[K/Fe] & \Huge[Ti/Fe] & \Huge[V/Fe] & \Huge[Cr/Fe] & \Huge[Mn/Fe] & \Huge[Ce/Fe]\\
 \hline 
 \hline     
 \\  
\Huge2M17140389-2929183 & \Huge0.18 &  & \Huge-0.43 & \Huge0.11 & \Huge0.17 & \Huge0.60 & \Huge0.24 & \Huge0.13 & \Huge-0.50 & \Huge-0.03 &  & \Huge1.00 & \Huge-1.28 & \Huge0.09 & \Huge0.44 & \Huge-0.11 & \Huge-0.15 & \Huge-0.46 & \Huge-0.12 & \Huge0.09 \\[2pt]
  
\Huge2M17141615-2925204 & \Huge0.22 &  & \Huge0.08 & \Huge0.24 & \Huge0.24 & \Huge0.52 & \Huge0.25 & \Huge0.16 & \Huge-0.47 & \Huge0.04 &  & \Huge0.11 & \Huge-0.37 & \Huge0.32 & \Huge0.44 & \Huge0.20 & \Huge-0.43 & \Huge-1.01 & \Huge-0.28 & \Huge-0.05 \\[2pt]

\Huge2M17141646-2926455 & \Huge0.20 &  & \Huge0.06 & \Huge0.14 & \Huge0.22 & \Huge0.40 & \Huge0.19 & \Huge0.18 & \Huge-0.53 & \Huge0.12 &  & \Huge0.47 & \Huge {...} & \Huge0.32 & \Huge0.50 & \Huge0.11 & \Huge0.42 & \Huge-0.40 & \Huge-0.30 & \Huge0.03 \\[2pt]

\Huge2M17142210-2930465 & \Huge0.21 &  & \Huge-0.22 & \Huge0.22 & \Huge0.23 & \Huge0.59 & \Huge0.19 & \Huge0.21 & \Huge-0.47 & \Huge0.07 &  & \Huge1.02 & \Huge0.67 & \Huge-0.05 & \Huge0.20 & \Huge0.12 & \Huge-0.06 & \Huge0.05 & \Huge-0.21 & \Huge0.23 \\[2pt]
  
\Huge2M17142243-2929316 & \Huge0.22 &  & \Huge-0.23 & \Huge0.23 & \Huge0.26 & \Huge0.55 & \Huge0.20 & \Huge0.22 & \Huge-0.46 & \Huge0.07 &  & \Huge1.13 & \Huge0.76 & \Huge0.42 & \Huge0.34 & \Huge0.38 & \Huge0.34 & \Huge0.00 & \Huge-0.17 & \Huge0.24 \\[2pt]

\Huge2M17142549-2926198 & \Huge0.24 &  & \Huge0.11 & \Huge0.26 & \Huge0.29 & \Huge0.34 & \Huge0.25 & \Huge0.17 & \Huge-0.35 & \Huge0.02 &  & \Huge0.18 & \Huge-0.06 & \Huge0.26 & \Huge0.27 & \Huge0.05 & \Huge-0.23 & \Huge-0.07 & \Huge-0.26 & \Huge0.05 \\[2pt]

\Huge2M17142716-2929529 & \Huge0.22 &  & \Huge-0.30 & \Huge0.16 & \Huge0.26 & \Huge0.59 & \Huge0.17 & \Huge0.24 & \Huge-0.49 & \Huge0.08 &  & \Huge1.09 & \Huge0.42 & \Huge0.28 & \Huge0.37 & \Huge0.37 & \Huge0.49 & \Huge-0.09 & \Huge-0.22 & \Huge0.08 \\[2pt]
  
\Huge2M17142865-2927003 & \Huge0.24 &  & \Huge-0.44 & \Huge0.23 & \Huge0.30 & \Huge0.65 & \Huge0.17 & \Huge0.24 & \Huge-0.48 & \Huge0.12 &  & \Huge1.19 & \Huge0.54 & \Huge0.30 & \Huge0.53 & \Huge0.48 & \Huge0.33 & \Huge0.06 & \Huge-0.17 & \Huge0.32 \\[2pt]

\Huge2M17142941-2922414 & \Huge0.23 &  & \Huge-0.22 & \Huge0.18 & \Huge0.25 & \Huge0.65 & \Huge0.21 & \Huge0.23 & \Huge-0.46 & \Huge0.07 &  & \Huge1.13 & \Huge0.28 & \Huge0.36 & \Huge0.40 & \Huge0.33 & \Huge0.17 & \Huge0.04 & \Huge-0.16 & \Huge0.33 \\[2pt]

\Huge2M17143081-2927457 & \Huge0.22 &  & \Huge0.18 & \Huge0.28 & \Huge0.29 & \Huge0.36 & \Huge0.22 & \Huge0.16 & \Huge-0.51 & \Huge0.04 &  & \Huge0.17 & \Huge0.12 & \Huge0.53 & \Huge0.23 & \Huge {...} & \Huge0.01 & \Huge-0.01 & \Huge {...} & \Huge {...} \\[2pt]

\Huge2M17143117-2930149 & \Huge0.21 &  & \Huge0.14 & \Huge0.24 & \Huge0.23 & \Huge0.33 & \Huge0.25 & \Huge0.15 & \Huge-0.43 & \Huge0.02 &  & \Huge0.12 & \Huge0.02 & \Huge0.45 & \Huge0.30 & \Huge0.12 & \Huge0.10 & \Huge-0.12 & \Huge-0.24 & \Huge-0.23 \\[2pt]

\Huge2M17143158-2932145& \Huge0.21 &  & \Huge0.14 & \Huge0.26 & \Huge0.25 & \Huge0.43 & \Huge0.25 & \Huge0.14 & \Huge-0.48 & \Huge0.01 &  & \Huge0.46 & \Huge-0.17 & \Huge0.18 & \Huge0.22 & \Huge0.09 & \Huge0.84 & \Huge-0.78 & \Huge-0.22 & \Huge-0.09 \\[2pt]
  
\Huge2M17143280-2927477 & \Huge0.20 &  & \Huge0.11 & \Huge0.27 & \Huge0.28 & \Huge0.29 & \Huge0.20 & \Huge0.13 & \Huge-0.53 & \Huge0.02 &  & \Huge0.76 & \Huge0.27 & \Huge0.30 & \Huge0.21 & \Huge {...} & \Huge0.07 & \Huge0.01 & \Huge {...} & \Huge {...} \\[2pt]

\Huge2M17143323-2925484 & \Huge0.09 &  & \Huge-0.19 & \Huge0.06 & \Huge0.07 & \Huge {...} & \Huge0.11 & \Huge0.10 & \Huge-0.54 & \Huge0.09 &  & \Huge0.82 & \Huge {...} & \Huge {...} & \Huge {...} & \Huge {...} & \Huge0.27 & \Huge {...} & \Huge {...} & \Huge {...} \\[2pt]

\Huge2M17143528-2929251 & \Huge0.23 &  & \Huge-0.25 & \Huge0.20 & \Huge0.25 & \Huge0.55 & \Huge0.21 & \Huge0.23 & \Huge-0.45 & \Huge0.05 &  & \Huge1.06 & \Huge0.72 & \Huge0.27 & \Huge0.33 & \Huge0.35 & \Huge0.28 & \Huge0.14 & \Huge-0.22 & \Huge0.23 \\[2pt]

\Huge2M17143573-2927050 & \Huge0.14 &  & \Huge-0.28 & \Huge0.15 & \Huge0.12 & \Huge0.17 & \Huge0.12 & \Huge0.17 & \Huge-0.65 & \Huge0.12 &  & \Huge1.00 & \Huge0.46 & \Huge0.28 & \Huge0.24 & \Huge0.40 & \Huge0.32 & \Huge0.05 & \Huge-0.07 & \Huge {...} \\[2pt]
  
\Huge2M17143640-2928513 & \Huge0.22 &  & \Huge0.15 & \Huge0.27 & \Huge0.30 & \Huge0.35 & \Huge0.21 & \Huge0.14 & \Huge-0.46 & \Huge0.08 &  & \Huge0.17 & \Huge0.23 & \Huge0.33 & \Huge0.21 & \Huge0.22 & \Huge-0.10 & \Huge-0.03 & \Huge-0.17 & \Huge-0.20 \\[2pt]

\Huge2M17143930-2926411 & \Huge0.19 &  & \Huge0.09 & \Huge0.24 & \Huge0.23 & \Huge0.33 & \Huge0.22 & \Huge0.12 & \Huge-0.41 & \Huge0.04 &  & \Huge0.17 & \Huge0.07 & \Huge0.31 & \Huge0.24 & \Huge0.21 & \Huge-0.11 & \Huge-0.01 & \Huge-0.21 & \Huge-0.23 \\[2pt]

\Huge2M17143986-2924168 & \Huge0.22 &  & \Huge-0.43 & \Huge0.10 & \Huge0.23 & \Huge0.44 & \Huge0.23 & \Huge0.19 & \Huge-0.49 & \Huge0.03 &  & \Huge1.09 & \Huge0.25 & \Huge0.36 & \Huge0.33 & \Huge0.14 & \Huge-1.51 & \Huge-0.40 & \Huge-0.20 & \Huge0.06 \\[2pt]
   
\Huge2M17144355-2929202 & \Huge0.19 &  & \Huge-0.30 & \Huge0.16 & \Huge0.15 & \Huge0.53 & \Huge0.17 & \Huge0.26 & \Huge-0.47 & \Huge0.04 &  & \Huge1.06 & \Huge0.75 & \Huge0.14 & \Huge0.36 & \Huge0.39 & \Huge0.24 & \Huge0.14 & \Huge-0.17 & \Huge0.15 \\[2pt]

\Huge2M17144517-2923296 & \Huge0.23 &  & \Huge0.15 & \Huge0.27 & \Huge0.23 & \Huge0.40 & \Huge0.24 & \Huge0.22 & \Huge-0.55 & \Huge0.05 &  & \Huge0.11 & \Huge0.21 & \Huge0.32 & \Huge0.36 & \Huge0.13 & \Huge0.05 & \Huge-0.19 & \Huge-0.21 & \Huge-0.23 \\[2pt]

\Huge2M17144561-2925135 & \Huge0.22 &  & \Huge0.12 & \Huge0.26 & \Huge0.26 & \Huge0.33 & \Huge0.24 & \Huge0.18 & \Huge-0.47 & \Huge-0.03 &  & \Huge0.15 & \Huge-0.20 & \Huge0.52 & \Huge0.30 & \Huge0.19 & \Huge0.09 & \Huge-0.54 & \Huge-0.18 & \Huge-0.14 \\[2pt]

\Huge2M17144732-2927090 & \Huge0.20 &  & \Huge0.10 & \Huge0.33 & \Huge0.24 & \Huge0.35 & \Huge0.21 & \Huge0.15 & \Huge-0.51 & \Huge0.04 &  & \Huge0.12 & \Huge0.40 & \Huge0.27 & \Huge0.20 & \Huge-0.03 & \Huge-0.77 & \Huge-1.13 & \Huge-0.31 & \Huge-0.04 \\[2pt]

\Huge2M17144851-2930108 & \Huge0.19 &  & \Huge-0.42 & \Huge-0.05 & \Huge0.19 & \Huge0.35 & \Huge0.25 & \Huge0.12 & \Huge-0.48 & \Huge-0.04 &  & \Huge1.04 & \Huge-1.29 & \Huge0.40 & \Huge0.19 & \Huge0.13 & \Huge-0.24 & \Huge-0.59 & \Huge-0.20 & \Huge0.19 \\[2pt]

\Huge2M17144988-2928301 & \Huge0.17 &  & \Huge-0.09 & \Huge0.16 & \Huge0.18 & \Huge0.43 & \Huge 0.15& \Huge0.18 & \Huge-0.52 & \Huge0.04 &  & \Huge0.75 & \Huge0.50 & \Huge0.41 & \Huge0.33 & \Huge0.25 & \Huge0.30 & \Huge-0.16 & \Huge-0.27 & \Huge0.05 \\[2pt]

\Huge2M17145331-2925511 & \Huge0.24 &  & \Huge-0.63 & \Huge-0.13 & \Huge0.22 & \Huge0.53 & \Huge0.25 & \Huge0.25 & \Huge-0.52 & \Huge0.04 &  & \Huge1.22 & \Huge0.27 & \Huge0.35 & \Huge0.47 & \Huge0.32 & \Huge-0.51 & \Huge-1.74 & \Huge-0.17 & \Huge0.05 \\[2pt]

\Huge2M17145815-2927517 & \Huge0.24 &  & \Huge-0.28 & \Huge0.18 & \Huge0.24 & \Huge0.54 & \Huge0.18 & \Huge0.29 & \Huge-0.44 & \Huge-0.03 &  & \Huge1.09 & \Huge0.41 & \Huge0.11 & \Huge0.41 & \Huge0.31 & \Huge0.21 & \Huge-0.26 & \Huge-0.18 & \Huge0.16 \\[2pt]
\\
\hline
\Huge \textbf{Cluster} & \Huge0.21$\pm0.01$ &  & \Huge-0.12$\pm0.05$ & \Huge0.19$\pm0.02$ & \Huge0.23$\pm0.01$ & \Huge0.45$\pm0.02$ & \Huge0.21$\pm0.01$ & \Huge0.18$\pm0.01$ & \Huge-0.49$\pm0.01$ & \Huge0.04$\pm0.01$ &  & \Huge0.69$\pm0.08$ & \Huge0.16$\pm0.11$ & \Huge0.30$\pm0.03$ & \Huge0.33$\pm0.02$ & \Huge0.21$\pm0.03$ & \Huge0.02$\pm0.09$ & \Huge-0.29$\pm0.09$ & \Huge-0.2$\pm0.01$ & \Huge0.05$\pm0.04$\\[8pt]
 \hline
 \hline 
 \end{tabular}
}
\end{table}

\section{Errors for NGC 6304}

\begin{longtable}{lcccccccc}
\caption{\label{tab:D1} Estimated errors in abundances, due to errors on atmospheric parameters and spectral noise, compared with the observed errors.} \\[4pt]
\hline
\hline 
ID & $\Delta T_{\mathrm{eff}}$ & $\Delta \log(g)$ & $\Delta \xi_{\mathrm{t}}$ & $\Delta [Fe/H]$ & $\sigma_{\mathrm{S/N}}$ & $\sigma_{\mathrm{tot}}$ & $\sigma_{\mathrm{obs}}$ & $\sigma_{\mathrm{ASPCAP}}$ \\[3pt]
\hline    
\hline
\endfirsthead

\caption[]{continued.} \\
\hline
\hline
ID &  $\Delta T_{\mathrm{eff}}$ &   $\Delta \log(g)$ &   $\Delta \xi_{\mathrm{t}}$ &   $\Delta [Fe/H]$ &   $\sigma_{\mathrm{S/N}}$ &   $\sigma_{\mathrm{tot}}$ &   $\sigma_{\mathrm{obs}}$ &   $\sigma_{\mathrm{ASPCAP}}$ \\[3pt]
\hline
\hline
\endhead

\multicolumn{9}{l}{\textbf{  Spectroscopy}} \\[3pt]
\hline
\hline

  $\Delta([C/Fe])$ &  0.05 &  0.09 &  0.04 &  0.04 &  0.08 &  0.14 &  0.28$\pm$ 0.03 &  0.03 \\[4pt]

  $\Delta([N/Fe])$ &  0.05 &  0.11 &  0.09 &  0.07 &  0.03 &  0.17 &  0.54$\pm$ 0.03 &  0.03 \\[4pt]

  $\Delta([O/Fe])$ &  0.11 &  0.02 &  0.03 &  0.03 &  0.04 &  0.13 &  0.11$\pm$ 0.01 &  0.03 \\[4pt]
  
  $\Delta([Na/Fe])$ &  0.03 &  0.06 &  0.02 &  0.02 &  0.02 &  0.08 &  0.20$\pm$ 0.05 &  0.13 \\[4pt]
  
  $\Delta([Mg/Fe])$ &  0.03 &  0.05 &  0.02 &  0.03 &  0.07 &  0.10 &  0.09$\pm$ 0.01 &  0.02 \\[4pt]

  $\Delta([Al/Fe])$ &  0.09 &  0.04 &  0.06 &  0.02 &  0.05 &  0.13 &  0.12$\pm$ 0.01 &  0.04 \\[4pt]

  $\Delta([Si/Fe])$ &  0.05 &  0.05 &  0.04 &  0.05 &  0.04 &  0.10 &  0.07$\pm$ 0.01 &  0.03 \\[4pt]

  $\Delta([S/Fe])$ &  0.04 &  0.11 &  0.08 &  0.10 &  0.09 &  0.20 &  0.32$\pm$ 0.04 &  0.09 \\[4pt]

  $\Delta([K/Fe])$ &  0.11 &  0.04 &  0.08 &  0.14 &  0.08 &  0.21 &  0.18$\pm$ 0.03 &  0.06 \\[4pt]

  $\Delta([Ca/Fe])$ &  0.07 &  0.04 &  0.04 &  0.04 &  0.06 &  0.12 &  0.09$\pm$ 0.01 &  0.03 \\[4pt]
  
  $\Delta([Ti/Fe])$ &  0.07 &  0.05 &  0.04 &  0.05 &  0.07 &  0.13 &  0.29$\pm$ 0.03 &  0.04 \\[4pt]

  $\Delta([V/Fe])$ &  0.13 &  0.05 &  0.05 &  0.04 &  0.04 &  0.16 &  0.13$\pm$ 0.03 &  0.14\\[4pt]

  $\Delta([Cr/Fe])$ &  0.03 &  0.06 &  0.06 &  0.07 &  0.06 &  0.13 &  0.23$\pm$ 0.04 &  0.08 \\[4pt]

  $\Delta([Mn/Fe])$ &  0.04 &  0.06 &  0.03 &  0.04 &  0.04 &  0.10 &  0.16$\pm$ 0.02 &  0.03 \\[4pt]

  $\Delta[Fe/H])$ &  0.04 &  0.04 &  0.01 &  0.01 &  0.02 &  0.06 &  0.05$\pm$ 0.00 &  0.01 \\[4pt]
  
  $\Delta([Ni/Fe])$ &  0.04 &  0.07 &  0.02 &  0.03 &  0.03 &  0.09 &  0.08$\pm$ 0.01 &  0.03 \\[4pt]

  $\Delta([Ce/Fe])$ &  0.04 &  0.01 &  0.03 &  0.03 &  0.07 &  0.09 &  0.10$\pm$  0.02 &  0.13 \\[4pt]
\hline       
\hline \\ [4pt]
\multicolumn{9}{l}{\textbf{Photometry}} \\[3pt]
\hline
\hline
$\Delta([C/Fe])$ & 0.08 & 0.05 & 0.09 & 0.07 & 0.07 & 0.16 & 0.23$\pm$0.03 & 0.03 \\[4pt]

$\Delta([N/Fe])$ & 0.07 & 0.09 & 0.08 & 0.08 & 0.03 & 0.16 & 0.46$\pm$0.03 & 0.03 \\[4pt]

$\Delta([O/Fe])$ & 0.10 & 0.05 & 0.06 & 0.06 & 0.04 & 0.15 & 0.11$\pm$0.01 & 0.03 \\[4pt]
  
$\Delta([Na/Fe])$ & 0.05 & 0.02 & 0.02 & 0.01 & 0.00 & 0.06 & 0.17$\pm$0.04 & 0.13 \\[4pt]
  
$\Delta([Mg/Fe])$ & 0.02 & 0.05 & 0.05 & 0.02 & 0.10 & 0.13 & 0.11$\pm$0.01 & 0.02 \\[4pt]

$\Delta([Al/Fe])$ & 0.05 & 0.05 & 0.10 & 0.06 & 0.09 & 0.16 & 0.14$\pm$0.02 & 0.04 \\[4pt]

$\Delta([Si/Fe])$ & 0.06 & 0.07 & 0.04 & 0.04 & 0.08 & 0.13 & 0.10$\pm$0.01 & 0.03  \\[4pt]

$\Delta([S/Fe])$ & 0.02 & 0.03 & 0.02 & 0.01 & 0.15 & 0.16 & 0.27$\pm$0.04 & 0.09\\[4pt]

$\Delta([K/Fe])$ & 0.03 & 0.04 & 0.03 & 0.03 & 0.11 & 0.13 & 0.19$\pm$0.03 & 0.06 \\[4pt]

$\Delta([Ca/Fe])$ & 0.07 & 0.03 & 0.06 & 0.05 & 0.09 & 0.14 & 0.09$\pm$0.01 &  0.03 \\[4pt]
  
$\Delta([Ti/Fe])$ & 0.07 & 0.07 & 0.01 & 0.02 & 0.19 & 0.22 & 0.27$\pm$0.03 & 0.04 \\[4pt]

$\Delta([V/Fe])$ & 0.12 & 0.06 & 0.04 & 0.02 & 0.01 & 0.14 & 0.17$\pm$0.04 & 0.14 \\[4pt]

$\Delta([Cr/Fe])$ & 0.02 & 0.03 & 0.07 & 0.01 & 0.14 & 0.16 & 0.20$\pm$0.04 & 0.08 \\[4pt]

$\Delta([Mn/Fe])$ & 0.02 & 0.03 & 0.01 & 0.01 & 0.08 & 0.09 & 0.13$\pm$0.02 & 0.03 \\[4pt]

$\Delta[Fe/H])$ & 0.03 & 0.04 & 0.05 & 0.02 & 0.03 & 0.08 & 0.08$\pm$0.00 & 0.01 \\[4pt]
  
$\Delta([Ni/Fe])$ & 0.05 & 0.01 & 0.02 & 0.05 & 0.07 & 0.10 & 0.07$\pm$0.01 & 0.03 \\[4pt]

$\Delta([Ce/Fe])$ & 0.04 & 0.06 & 0.02 & 0.01 & 0.06 & 0.10 & 0.16$\pm$0.03 & 0.13\\[4pt]
\hline 
\end{longtable}

\section{Spectral line quality classification} \label{Appendix E}

To assess the reliability of the abundance determinations, we classified the quality of the spectral lines used for each element based on three independent criteria, each normalized to a value between 0 and 1, where 1 represents the optimal case. The first criterion evaluated the line coverage across the sample and was defined as the ratio between the number of lines successfully used in the full sample (27 stars) and the total number of available lines for the element. The second criterion was based on the S/N of the lines. For each element, we compute the average S/N of its lines across all stars and normalize this value by an expected optimal S/N of 100. The third criterion considers the internal line-to-line scatter, defined as the ratio between the internal uncertainty reported by {\ttfamily BACCHUS} (i.e., the standard deviation among the lines used for each element) and the $\sigma_{\rm obs}$ across the sample. Next, we define a composite score, referred to as the "goodness value," as the average of the three criteria described before. This final value ranges from 0 to 1 and provides a global assessment of the quality of the abundance measurement for each element. Based on the goodness value, we classify each element into one of three quality categories:\\
\\
- {\ttfamily Good elements}: \textit{goodness value} $\geq$ 0.60\\
- {\ttfamily Intermediate elements}: 0.40 < \textit{goodness value} < 0.60\\
- {\ttfamily Bad elements}: \textit{goodness value} $\leq$ 0.40\\
\\
Based on our line-quality classification, we classified eight elements as \textit{Good} (C, O, Mg, Al, Si, Ca, Fe, and Ni), nine as \textit{Intermediate} (N, Na, S, K, Ti, V, Cr, Mn, and Ce), and three as \textit{Bad} (P, Co, and Nd). The latter elements were excluded from our analysis because of their unreliable measurements. This classification allows us to objectively identify which elements provide the most reliable abundance estimates in our sample. Table \ref{tab:D1} lists our quality estimation per element. We note that this assessment is in general good agreement with previous such APOGEE line assessments (e.g., \citealp{2018AJ....156..126J}).


\begin{table}[h]
\centering
\caption{Quality classification of the lines per element.}
\label{tab:E1}
\begin{minipage}{0.48\textwidth}
\centering
\begin{tabular}{lcc}
\hline
\hline
Elements & Goodness value & Classification\\
\hline
C & 0.60 & Good\\[2pt]
O & 0.70 & Good\\[2pt]
Mg & 0.81 & Good\\[2pt]
Al & 0.60 & Good\\[2pt]
Si & 0.80 & Good\\[2pt]
Ca & 0.72 & Good\\[2pt]
Fe & 0.97 & Good\\[2pt]
Ni & 0.71 & Good\\[2pt]
N & 0.56 & Intermediate\\[2pt]
Na & 0.42 & Intermediate\\[2pt]
\hline
\hline
\end{tabular}
\end{minipage}
\hfill
\begin{minipage}{0.4\textwidth}
\centering
\begin{tabular}{lcc}
\hline
\hline
Elements & Goodness value & Classification\\
\hline
S & 0.45 & Intermediate\\[2pt]
K & 0.59 & Intermediate\\[2pt]
Ti & 0.46 & Intermediate\\[2pt]
V & 0.49 & Intermediate\\[2pt]
Cr & 0.44 & Intermediate\\[2pt]
Mn & 0.45 & Intermediate\\[2pt]
Ce & 0.59 & Intermediate\\[2pt]
P & 0.35 & Bad\\[2pt]
Co & 0.33 & Bad\\[2pt]
Nd & 0.28 & Bad\\[2pt]
\hline
\hline
\end{tabular}
\end{minipage}
\end{table}

\section{{\ttfamily ASPCAP} and {\ttfamily BACCHUS} fit comparison for Al I and Si I lines in NGC 6304}

\begin{figure}
\centering
\includegraphics[width=0.9\textwidth, height=0.48\textheight]{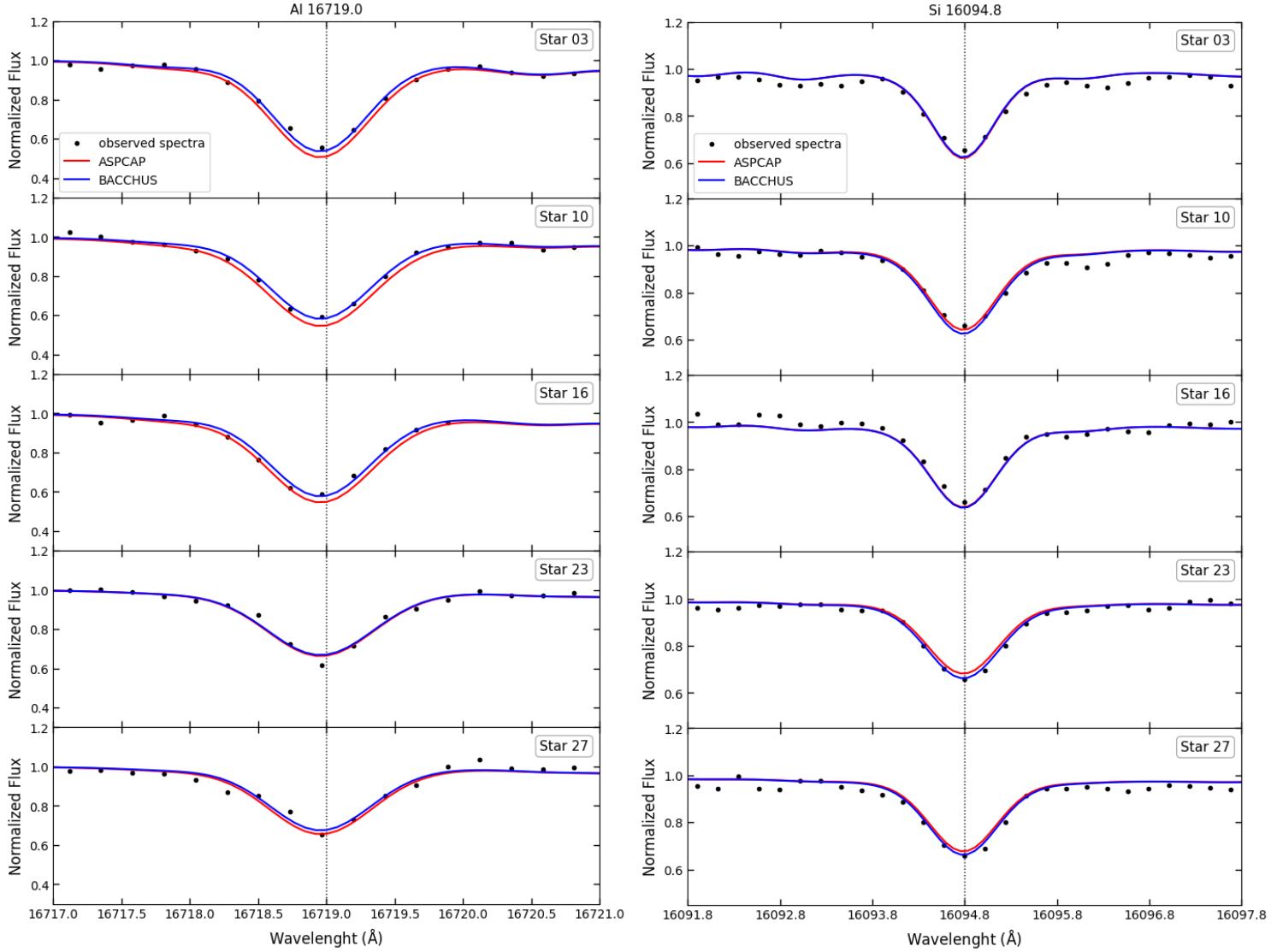}
\caption{Comparison between the observed APOGEE spectra (black dots) and synthetic spectra computed with {\ttfamily BACCHUS} using two different abundances for five representative stars in NGC 6304. The red lines correspond to synthetic spectra generated with the log(Si I) (or log(Al I)) abundance derived by {\ttfamily ASPCAP}, while the blue lines correspond to spectra computed using the {\ttfamily BACCHUS} abundance for the same element. The left panel shows Al I 16719.0 \AA, and the right panel shows Si I 16094.8 \AA. In most cases, the {\ttfamily BACCHUS} abundance results in a slightly better fit to the observed line profile, taking into account that both synthetic spectra were computed with the same radiative transfer code (Turbospectrum) and line list.}
\label{F1}
\end{figure}

In the panels g) and h) of Figure \ref{fig:8}, we observed an offset between the Si and Al abundances of {\ttfamily ASPCAP} compared to those derived with {\ttfamily BACCHUS} using the same stellar parameters. Then, to address this behavior, we assess the quality of the line fits produced by {\ttfamily BACCHUS} compared to those provided by ASPCAP. For that, we performed a visual inspection of individual spectral lines for five representative stars, as we see in Figure \ref{F1}.\\
Due to the lack of access to the internal synthetic spectra generated by {\ttfamily ASPCAP}, we adopted the following strategy: for each star, we generated two synthetic spectra using the {\ttfamily BACCHUS} code. The first synthetic spectrum was computed using the abundance of log(Si I) (or log(Al I)) reported by {\ttfamily ASPCAP}, while the second was generated using the abundance derived by {\ttfamily BACCHUS}. Then both synthetic spectra were compared with the observed APOGEE spectrum for the respective star. Figure \ref{F1} displays the observed spectra (black dots) along with the synthetic spectra computed with the {\ttfamily ASPCAP} abundances (red lines) and the {\ttfamily BACCHUS} abundances (blue lines) for the five selected stars. In most cases, the synthetic spectra based on the {\ttfamily BACCHUS} abundances provide a slightly better fit to the observed line profile than those based on {\ttfamily ASPCAP} values. This suggests that the {\ttfamily BACCHUS} analysis may produce more accurate individual line fits.\\
It is important to emphasize that the comparison does not directly evaluate the internal fit procedure of {\ttfamily ASPCAP}, but rather the agreement between its reported abundances and the observed line profiles when modeled with {\ttfamily BACCHUS}. Thus, while this approach provides a useful visual comparison, it is limited by the fact that both synthetic spectra were generated using the same radiative transfer engine (Turbospectrum) and line list as implemented in {\ttfamily BACCHUS}. Although this comparison did not significantly affect abundance determinations or conclusions presented in the main analysis, it highlights subtle differences in fitting quality that are worth noting for future studies.
\end{appendix}

\end{document}